\documentclass[prb,a4paper,twocolumn,showpacs,floatfix]{revtex4-1}
\usepackage[utf8]{inputenc}
\usepackage[T1]{fontenc}
\usepackage{lmodern}
\usepackage{amssymb}
\usepackage{amsmath}
\usepackage{amsbsy}
\usepackage{bm}
\usepackage{esint}
\usepackage{graphicx}
\usepackage{color}
\newcommand{\I}{\mathrm{i}}
\newcommand{\E}{\mathrm{e}}
\begin{document}
\title{Topological changes of two-dimensional magnetic textures}
\author{Ricardo Gabriel Elías}
\author{Alberto D.\ Verga}\email{Alberto.Verga@univ-amu.fr}
\affiliation{Université d'Aix-Marseille, IM2NP-CNRS, Campus St.\ Jérôme, Case 142, 13397 Marseille, France}
\date{\today}
\begin{abstract}
We investigate the interaction of magnetic vortices and skyrmions with a spin-polarized current. In a square lattice, fixed classical spins and quantum itinerant electrons, evolve according to the coupled Landau-Lifshitz and Schrödinger equations. Changes in the topology occur at microscopic time and length scales, and are shown to be triggered by the nucleation of a nontrivial electron-spin structure at the vortex core.
\end{abstract}
\maketitle

%
\section{Introduction}

Itinerant magnetism is a fascinating state of matter where the interplay of short range coupling (exchange, spin-orbit, crystal anisotropy) and long range dipolar interactions lead to a variety of spatial orders. Experiments show in particular that magnetic structures with nontrivial topology naturally arise in nanosized ferromagnets and chiral metals. Vortices are present in confined geometries with closed magnetic flux lines, such as permalloy nanodots;\cite{Cowburn-1999vn,Shinjo-2000dz,Wachowiak-2002pl} the Dzyaloshinskii-Moriya spin-orbit coupling in magnetic metals with inversion asymmetry like bulk MnSi or Fe atomic films, favors helical ordering in the form of a skyrmion lattice.\cite{Muhlbauer-2009vn,Yu-2010fk,Heinze-2011fk} The existence of inhomogeneous metastable states in two-dimensional isotropic ferromagnets, distinct from the usual domains, was theoretically predicted by Belavin and Polyakov,\cite{Belavin-1975xw} who exhibited an asymptotically uniform solution with a reversed magnetization in the central region. Lattices of skyrmions were shown to be thermodynamically allowed in chiral magnets, within a range of applied magnetic fields.\cite{Bogdanov-1989fk} The topology of these magnetization fields can be characterized by their degree, or topological charge;\cite{Dubrovin-1985fk} the skyrmion configuration realizes a map between the plane (the ferromagnetic film) and the sphere (the directions of the magnetization vector); it has therefore an integer topological charge.\cite{Belavin-1975xw,Kosevich-1990cy} From a topological point of view, the isolated vortices observed in nanomagnets, are more exotic, since their topological charge is a half-integer.\cite{Tretiakov-2007fc} Vortices with an out-of-plane core magnetization, can be viewed as half skyrmions, sometimes called merons,\cite{Gross-1978kx,Senthil-2004kx} because only a half sphere is mapped. In disk magnets, their stability is ensured by the constraint of a tangent magnetization at the boundary that minimizes the dipolar magnetic energy.\cite{Usov-1993oq}

Interestingly, experiments reveal that these topological configurations can be manipulated not only by external magnetic fields but also using purely electric means, by a spin polarized current through the spin-transfer torque mechanism.\cite{Berger-1996bh,Slonczewski-1996lq} The polarity of a vortex core can be reversed by applying a short pulse of an in-plane magnetic field,\cite{Van-Waeyenberge-2006fk} or by a current.\cite{Yamada-2008eu} More recently, ultrafast switching of an uniform magnetization, with the temporary formation of a magnetic singularity, was achieved in experiments using laser pulses of circularly polarized light,\cite{Vahaplar-2009ys,Hertel-2009vn} a technique that can in principle also be effective in vortex switching via a topological inverse Faraday effect.\cite{Taguchi-2012uq}

%
\begin{figure*}[t]
\centering
\includegraphics[width=0.33\textwidth]{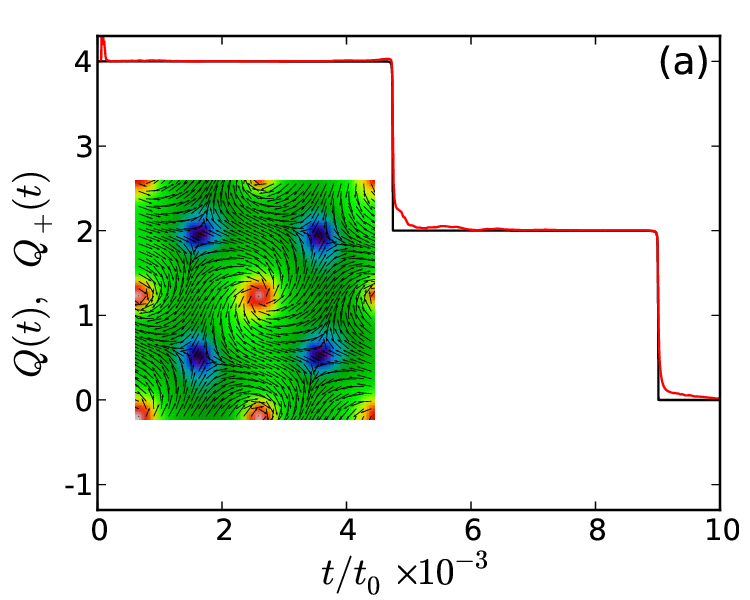}%
\includegraphics[width=0.33\textwidth]{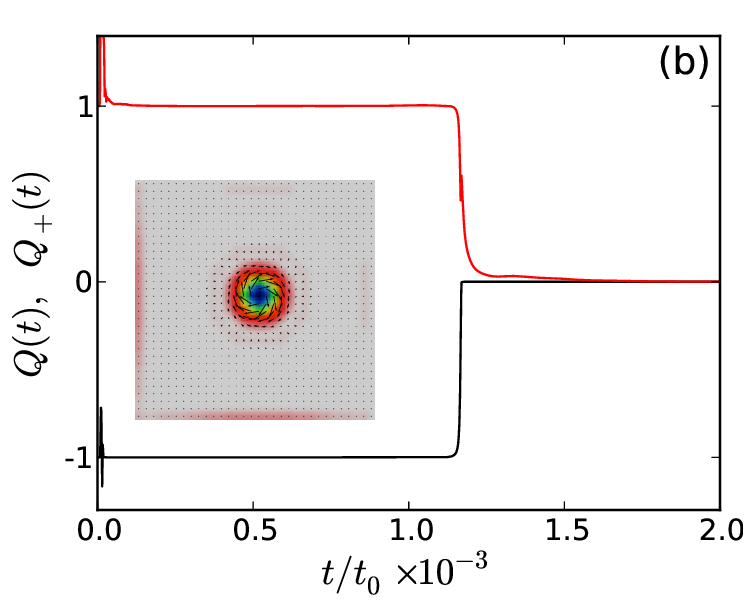}%
\includegraphics[width=0.33\textwidth]{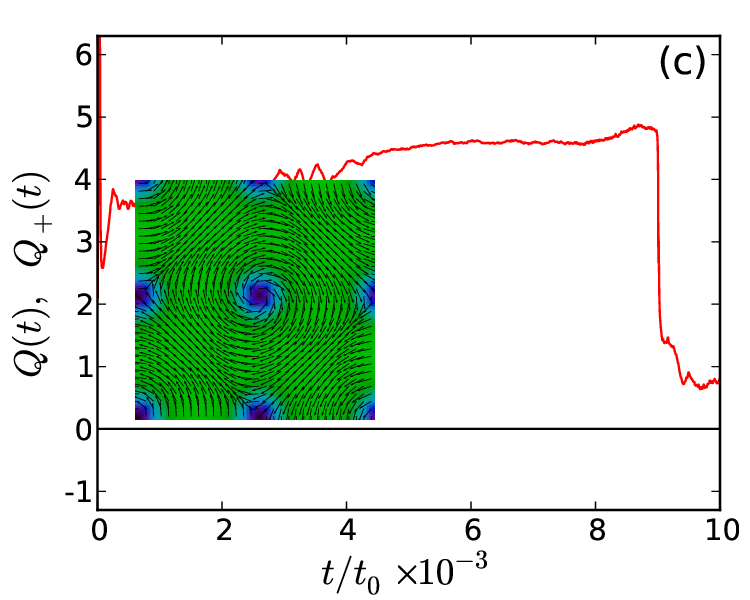}\\
\includegraphics[width=0.33\textwidth]{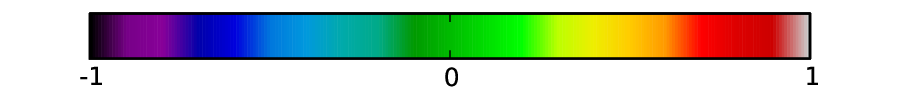}
\caption{\label{f:Qt} (Color online) Evolution of the topological charge $Q$ (black) and $Q_+$ (red), of a skyrmion lattice ($Q = 4$ per cell) with $x$-polarized electrons (a), a single $Q = -1$ skyrmion (b), and a vortex array (c), with $z$-polarized electrons. The insets show the initial magnetization field, arrows are for the $(S_x,S_y)$ components, and $S_z$ is in color, from black $S_z = -1$, to white $S_z = 1$ (bottom color bar). The lattice size is $L = 128\,a$ and the applied electric field $E/E_0=10^{-4}$; electron densities are $n_e=0.02$ for the skyrmion lattice (a), and $n_e=0.1$ for (b) and (c).}
\end{figure*}

The dynamics of the magnetization at the microscopic scale is governed by the Landau-Lifshitz equation,\cite{Landau-1935fk}
\begin{equation}\label{eq:LL}
\hbar\frac{\partial}{\partial t}\bm S=\bm S\times\bm f -
	\alpha \bm S\times(\bm S\times \bm f)\,,
\end{equation}
where $\bm S$ is the dimensionless spin (treated here as a classical variable proportional to the magnetization in an atomic volume $a^3$) and $\bm f$ the effective field derived from the free energy functional ($\hbar$ is the Planck constant, $\alpha$ is the damping constant, and $\bm f$ has the dimensions of energy). This equation takes into account the exact conservation of the magnetization norm to its saturation value (for convenience we define $|\bm S|=S=1$). In addition, the special mathematical form of (\ref{eq:LL}), where the right hand side is perpendicular to the spin vector, ensures the conservation of the topological charge,\cite{Papanicolaou-1991it} defined by,
\begin{equation}\label{eq:Q}
Q=\int_{\mathbb R^2} \frac{d\bm x }{4\pi}q(\bm x,t),
\quad q=\bm S\cdot\partial_x\bm S\times\partial_y\bm S\,,
\end{equation}
in a two-dimensional system, where $\bm x=(x,y)$ is a point in the plane perpendicular to $z$. It is worth noting that the conservation of the magnetization topology is independent of the effective field specific form, and it holds even in the presence of norm preserving dissipation and time-dependent external fields. However, the topology conservation is violated by stochastic perturbations, related for instance to thermal or quantum fluctuations.

The micromagnetic approach was extensively used in recent years to investigate the dynamics of magnetic textures involving monopoles and vortices. In spite of the fact that the Landau-Lifshitz equation conserves the topology of the magnetization field, micromagnetic simulations on discrete lattices proved to be useful in describing complex topological changes.\cite{Thiaville-2003rq} In particular, these simulations revealed the importance of the excitation of gyration modes and vortex-antivortex annihilation in vortex core reversal.\cite{Hertel-2006ib,Yamada-2007qt,Guslienko-2008so,Gaididei-2008vn,Weigand-2009la}

In this paper we investigate the topological changes of magnetic textures induced by a spin-polarized current, using a semiclassical two-dimensional lattice model, in which the itinerant electrons are quantum, and the fixed spins, classical. The interaction of the itinerant electrons with the fixed ones gives rise to a spin-transfer torque, generally described in the quasi-adiabatic limit by adding terms in the gradients of the magnetization,
\begin{equation}\label{eq:grads}
\bm v_s\cdot\nabla \bm S\,,
\end{equation}
where $\bm v_s$ is related to the electron spin polarized current.\cite{Bazaliy-1998qf,Zhang-2004ve} This approximation is not well suited in the presence of strong magnetization gradients, as it is precisely the case near vortex cores. In the present model we keep the full quantum electron dynamics to resolve nonlocal effects that are fundamental, as we will demonstrate, in the mechanisms involving the change of topology through the formation of magnetic singularities. As underlined by Miltat and Thiaville,\cite{Miltat-2002kx} the nucleation of Bloch points and vortex cores are at the edge of quantum magnetism.

%
\begin{figure*}[t]
\centering
\includegraphics[width=0.16\textwidth]{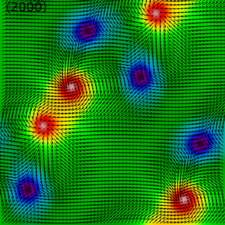}%
\includegraphics[width=0.16\textwidth]{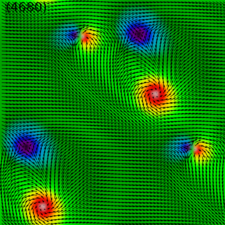}%
\includegraphics[width=0.16\textwidth]{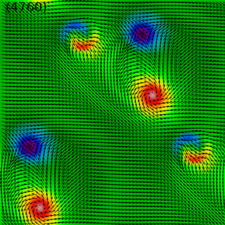}%
\includegraphics[width=0.16\textwidth]{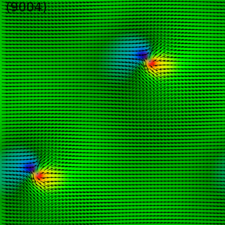}%
\includegraphics[width=0.16\textwidth]{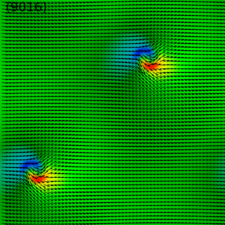}%
\includegraphics[width=0.16\textwidth]{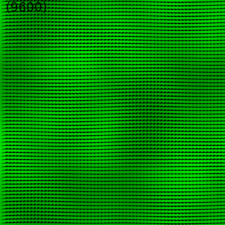}\\
\includegraphics[width=0.16\textwidth]{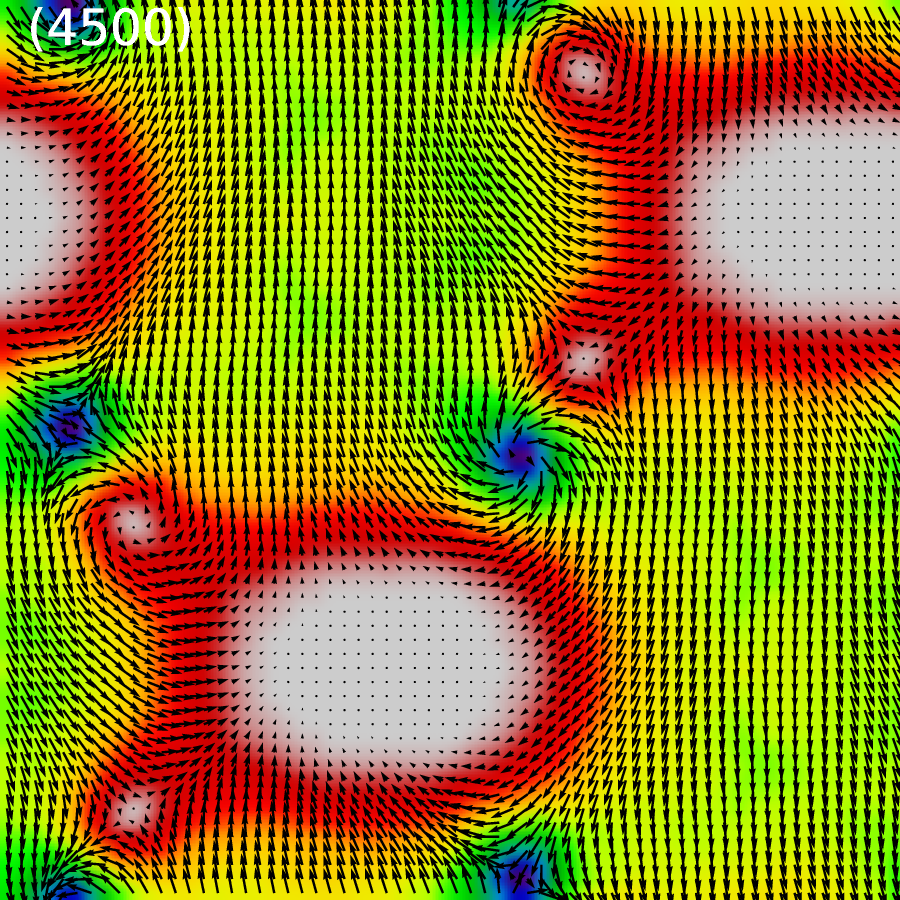}%
\includegraphics[width=0.16\textwidth]{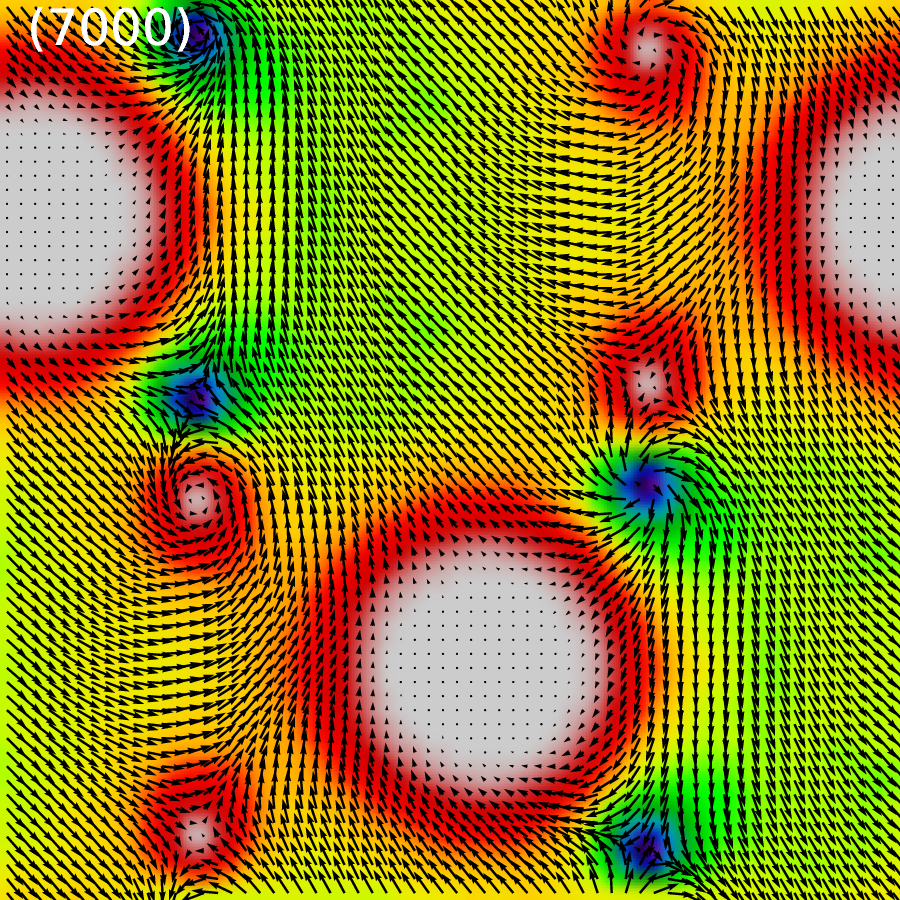}%
\includegraphics[width=0.16\textwidth]{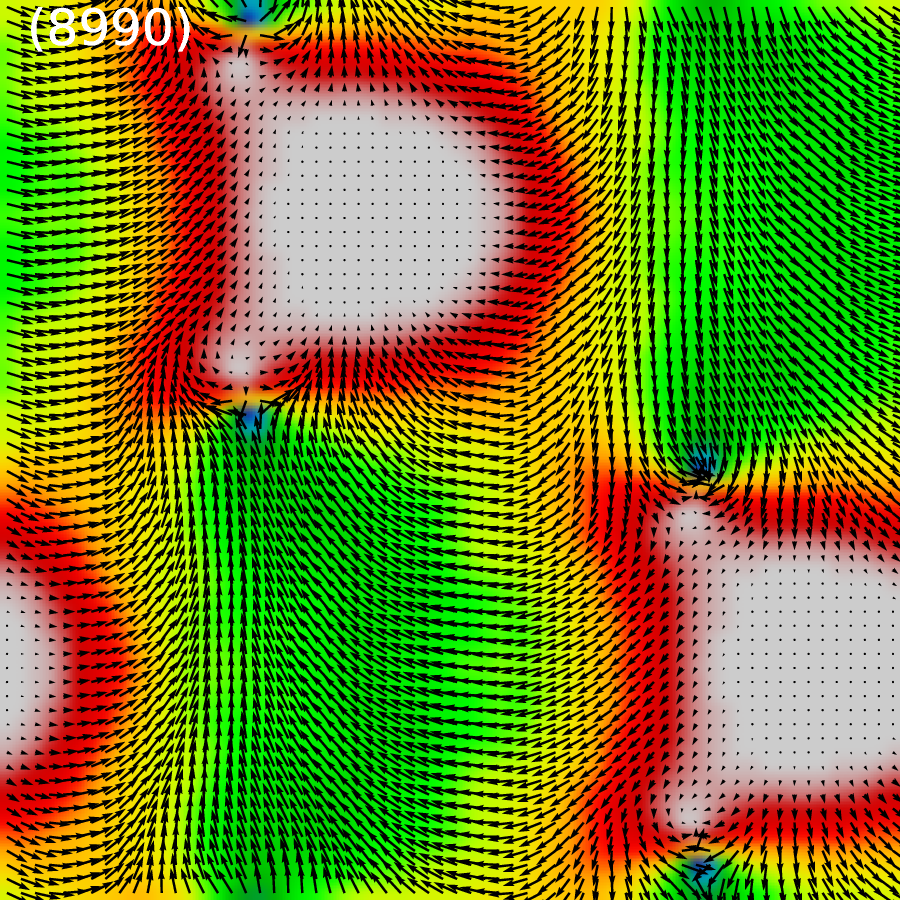}%
\includegraphics[width=0.16\textwidth]{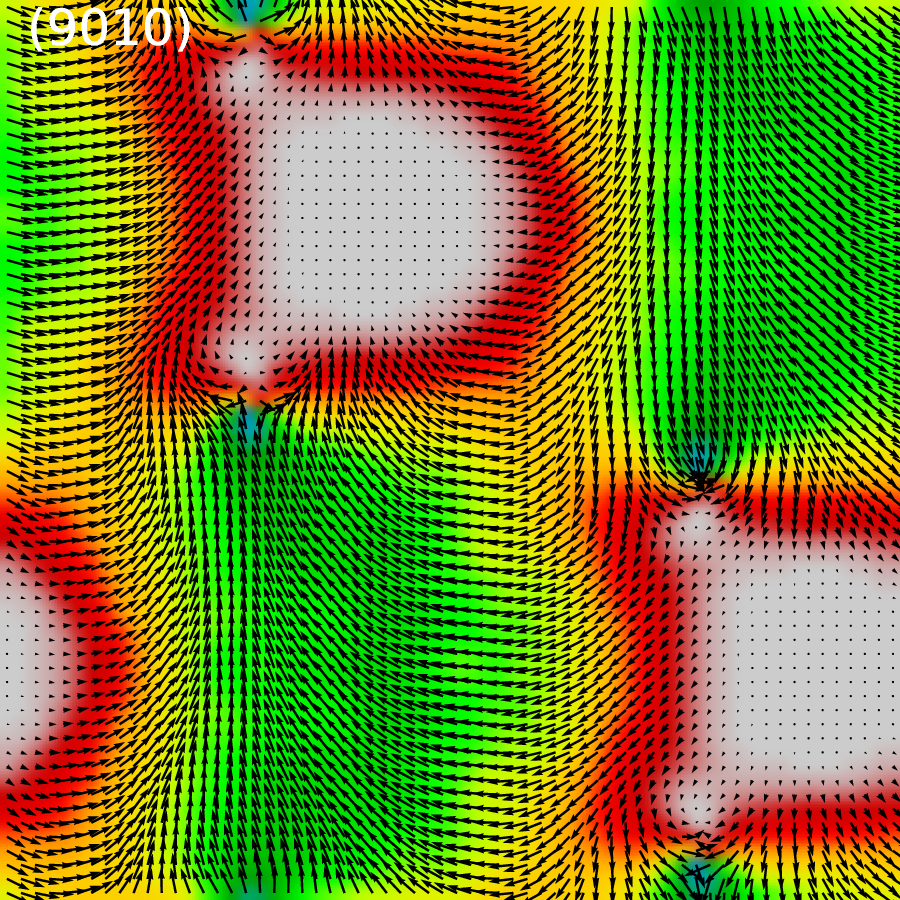}%
\includegraphics[width=0.16\textwidth]{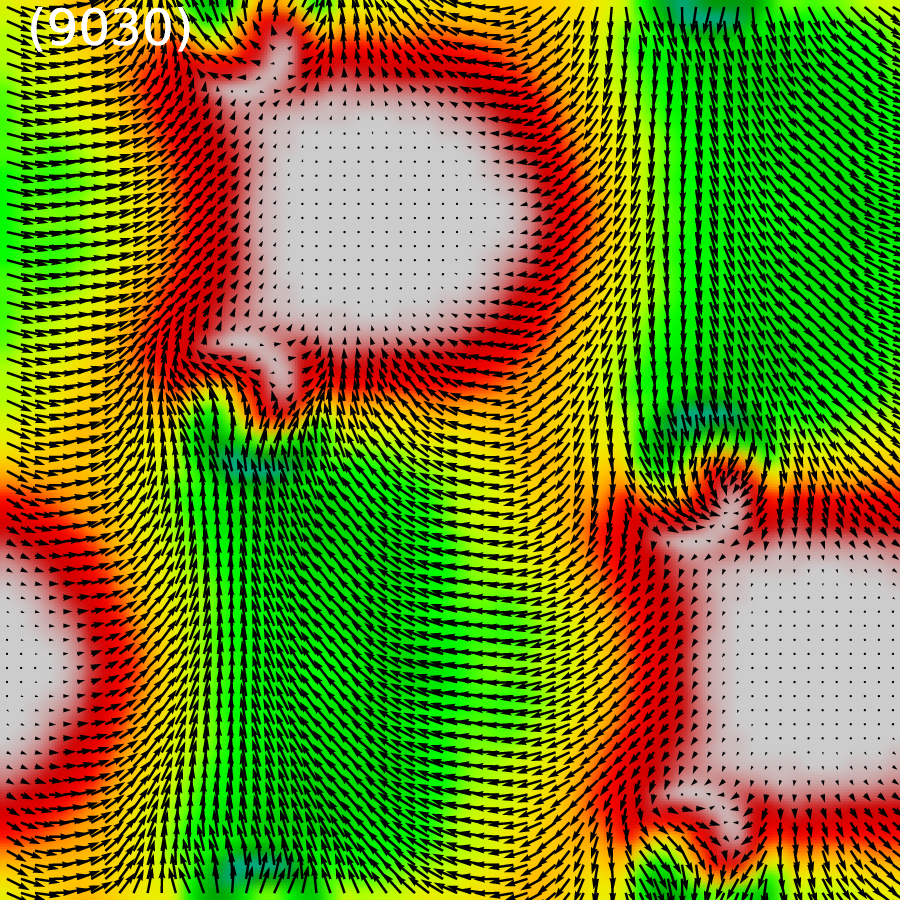}%
\includegraphics[width=0.16\textwidth]{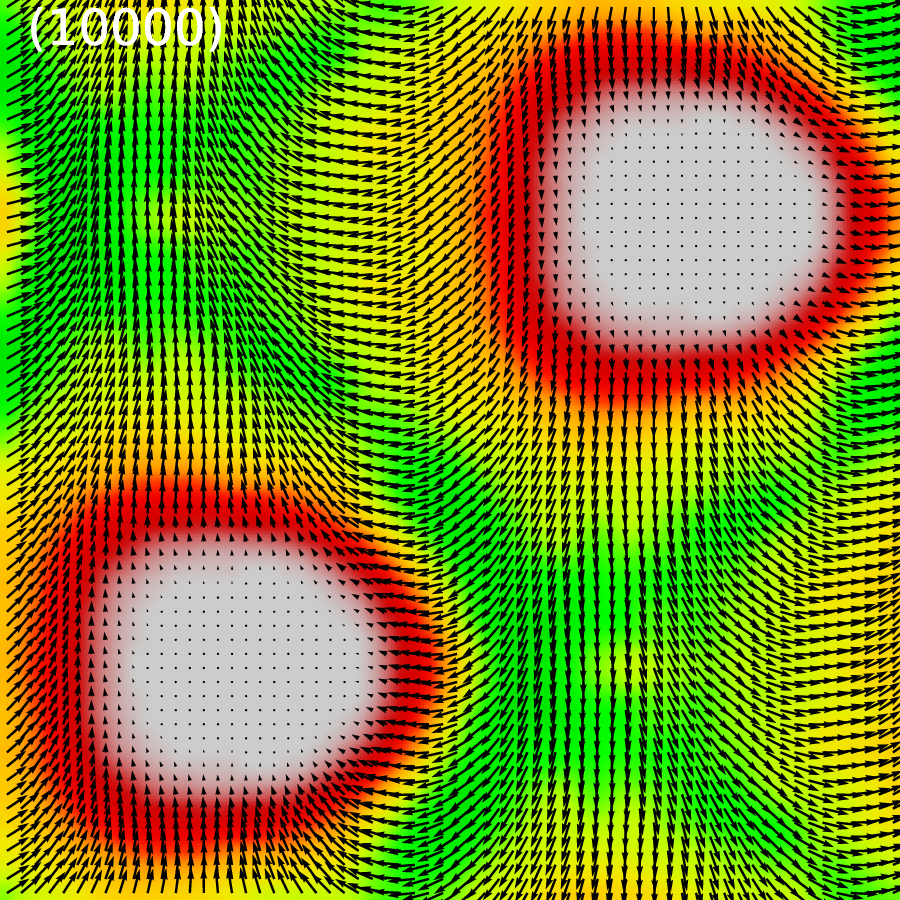}
\caption{\label{f:S} (Color online) Magnetization texture at selected times, for an initial skyrmion lattice (upper row) and vortex array (bottom row), corresponding to Fig.~\protect\ref{f:Qt} (a) and (c) respectively. Vortices interaction result in the change of the topological charge and the emission of spin waves.}
\end{figure*}

%
\section{Lattice model of coupled itinerant and fixed spins}

We consider a periodic square lattice of fixed spins $\bm S$ (classical, $|\bm S|=1$) and a single electron that can jump between neighboring sites $i=(x,y)=x\hat{x}+y\hat{y}$, and $j$ (the hopping energy is $\epsilon$, the electron charge $-e$, the lattice parameter $a$ and size $L$). Periodic boundary conditions ensure a well defined topology of the total system. A constant electric field $E\hat{x}$ is applied to create an electron current; this current is polarized by a fictive magnetic field acting only on the electrons $\bm B_p$. Electrons and fixed spins are coupled through an exchange interaction $J_s$, which in ferromagnets can be much larger than the Curie energy $J$.\cite{Tatara-2008kx} The electron Hamiltonian is,
\begin{equation}\label{eq:He}
H_e =  -\epsilon\sum_{\langle i,j\rangle} 
	\E^{\I \phi_{i,j}(t)}c_i^\dag c_j
	-J_s\sum_i \bm S_i \cdot (c_i^\dag \bm \sigma c_i)+H_p
\end{equation}
where $c_i=(c_{\uparrow i},c_{\downarrow i})$ is the annihilation operator at site $i$ and spin up $\uparrow$ or down $\downarrow$. To preserve the lattice periodicity we used a gauge transformation that introduces a time-dependent vector potential $\bm A=Et\hat{x}$, instead of a static electric potential. This allows us to take into account the constant electric field through the phase $\phi_{i,j}(t)=(i-j)\cdot\hat{x}eaEt/\hbar$, which is zero if the neighboring sites $i,j$ are not in the $x$ direction. The second term accounts for the interaction energy with the fixed spins ($\bm \sigma$ are the Pauli matrices). The last term, $H_p=-\mu_e\bm B_p \cdot \sum_i c_i^\dag \bm \sigma c_i$, allows the current to polarize in the direction $\bm B_p$ ($\mu_e$ is the electron magnetic moment). The magnetic energy is the sum of exchange $J>0$, anisotropy $K$ (positive or negative for  easy-plane or easy-axis cases, respectively), and Dzyaloshinskii-Moriya $D$ terms:
\begin{equation}\label{eq:HS}
H_S = \frac{J}{2} \sum_i (\nabla \bm S_i)^2
	+ \frac{K}{2} \sum_i  S_{zi}^2 - 
	\frac{D}{2}\sum_i\bm S_i\cdot (\nabla \times \bm S_i)
\end{equation}
where $\nabla$ is here the \emph{discrete} gradient operator (note that it is dimensionless). In the following we use units such that $a=\epsilon=\hbar=e=1$. The typical microscopic scales are $a\sim 0.3\,\mathrm{nm}$ and $\epsilon\sim 1\,\mathrm{eV}$ for a ferromagnet, or $a\sim 0.5\,\mathrm{nm}$ and $\epsilon\sim 0.1\,\mathrm{eV}$ for MnSi, given a time unit $t_0\sim 1-10\,\mathrm{fs}$, depending on the energy scale; the unit of electric field is about $E_0 \sim \epsilon/(ea) \approx 0.1-1\,10^9\,\mathrm{V\,m^{-1}}$ and the unit of current $I_0 \sim e\epsilon/\hbar\approx 10-100\,\mu\mathrm{A}$. These small time and length scales, related to the electron kinetic energy and lattice spacing, are necessary to track the changes in topology.

The system evolution is governed by the Schrödinger equation (or equivalently, the Heisenberg equation for the operators) for the electrons,
\begin{equation}\label{eq:c}
\I \dot{c}_i(t)=H_e(t,S_i) c_i(t)\,,
\end{equation}
and (\ref{eq:LL}) for the fixed spins, with $\bm f_i=-\partial H_S/\partial \bm S_i$,
\begin{equation}\label{eq:f}
\bm f_i= J\nabla^2 \bm S_i - K \bm S_{zi} + D \nabla \times \bm S_i+J_s n_e\bm s_i\,,
\end{equation}
where $\bm s_i=\langle c_i^\dag \bm \sigma c_i\rangle$ is the itinerant electron spin, and $\langle c_i^\dag  c_i\rangle=n_e$, the number of electrons per lattice site. The last term gives the spin-transfer torque due to the moving electrons. 

At variance to the linear response or quasi-adiabatic approximations, leading to a modified Landau-Lifshitz equation,\cite{Zhang-2004ve,Thiaville-2005cr,Tatara-2008kx} we keep the full electron dynamics (\ref{eq:c}) to compute the spin-transfer torque in (\ref{eq:f}) (see Ref.~\onlinecite{Ohe-2006db} for a related model). Indeed, the usual approach starts from the spin-current continuity equation (straightforwardly obtained from the Schrödinger equation):
\begin{equation}\label{eq:cont}
\frac{\partial \bm s}{\partial t}+\nabla\cdot {\cal J}=J_s \bm s\times\bm S-\Gamma,
\end{equation}
where ${\cal J}$ is the spin-current density tensor, and the last term $\Gamma$, absent in our model, takes into account the dissipation mechanisms. Eq.~(\ref{eq:cont}) is then solved by approximating the spin-transfer torque by a series in the gradients (\ref{eq:grads}), $\bm s \approx n_e \bm S + \delta {\bm s}$, and ${\cal J} \approx \bm v_s \otimes \bm S$, to obtain,\cite{Zhang-2004ve}
\begin{equation}\label{eq:zhang}
\delta \bm s \approx \bm S \times \bm v_s \cdot \nabla \bm S + \beta \bm v_s \cdot \nabla \bm S\,.
\end{equation}
Here, the first term\cite{Bazaliy-1998qf} gives the adiabatic contribution to the spin-transfer torque, and the second term, referred to as the ``beta'' term,\cite{Zhang-2004ve,Piechon-2007rm} is the non-adiabatic contribution coming from the $\Gamma$ spin relaxation in (\ref{eq:cont}). However, in the presence of large gradients and strongly non-stationary processes, characteristic of the topological transitions, the quasi-adiabatic approximation breaks down. In addition, the scattering of the itinerant electrons by the fixed spin inhomogeneities, might naturally lead to inhomogeneities in the electron current, also neglected in this approximation (\ref{eq:zhang}), which considers $\bm v_s$ as a constant. In our model the spin-current density $\cal J$ is a dynamical variable computed from the electron wave function. We will see that taking into account the itinerant electron quantum dynamics reveals mechanisms such as localization, scattering, and angular and linear momentum transfers that have an important role in the changes of the magnetization distribution (fixed spins).

%
\section{Current induced topological changes in magnetic textures}

%
\begin{figure*}
\centering
\includegraphics[width=0.24\textwidth]{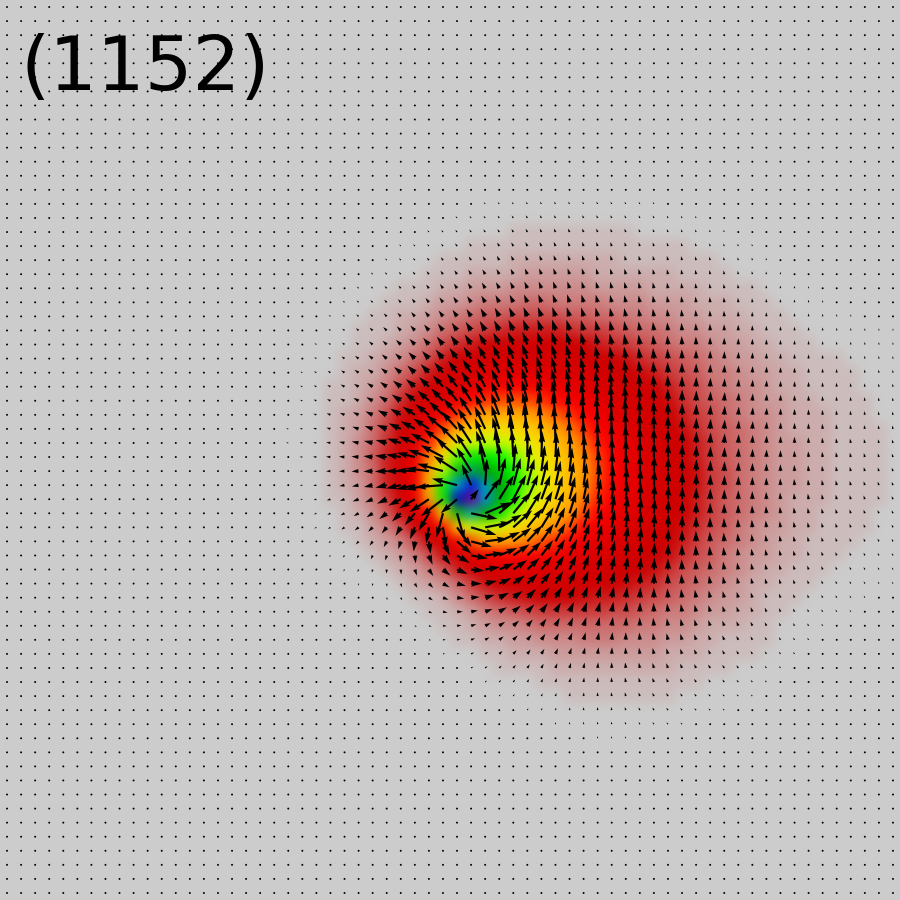}\hfill%
\includegraphics[width=0.24\textwidth]{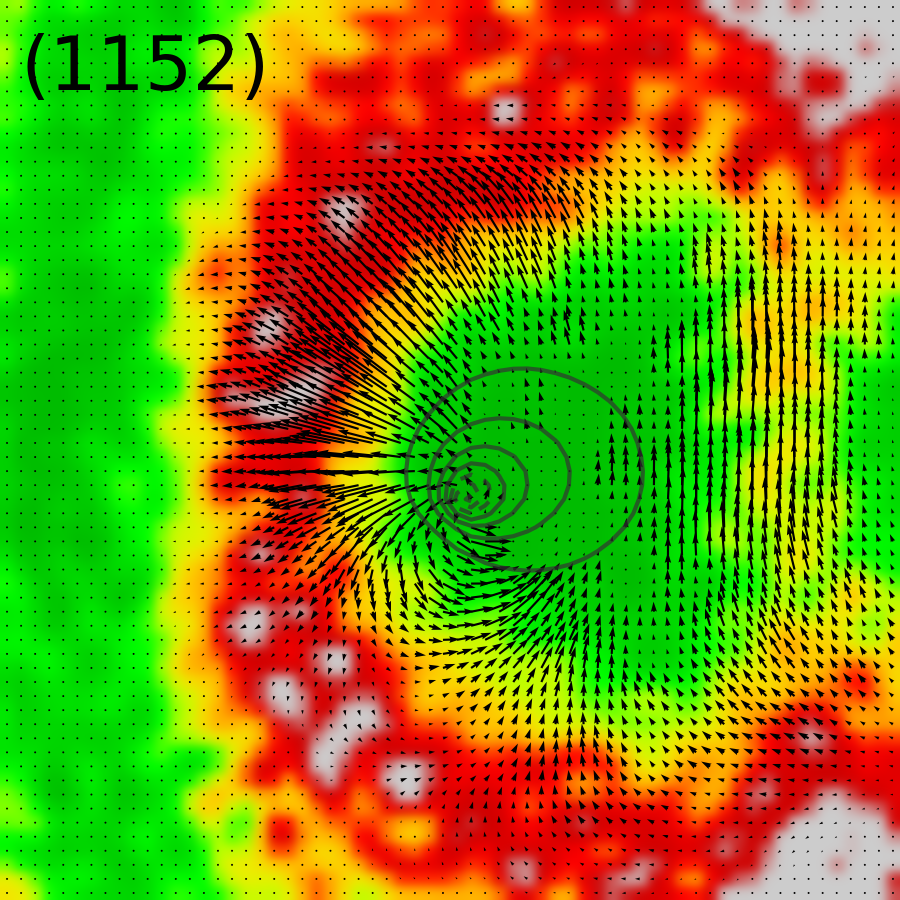}\hfill%
\includegraphics[width=0.24\textwidth]{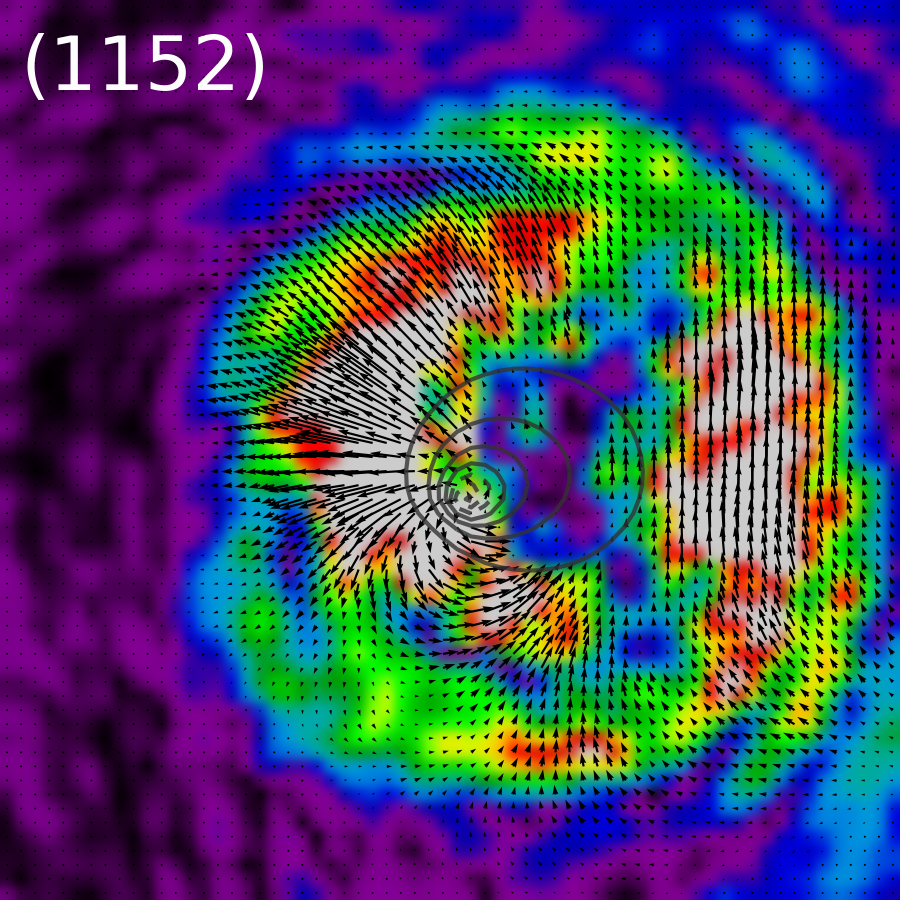}\hfill%
\includegraphics[width=0.24\textwidth]{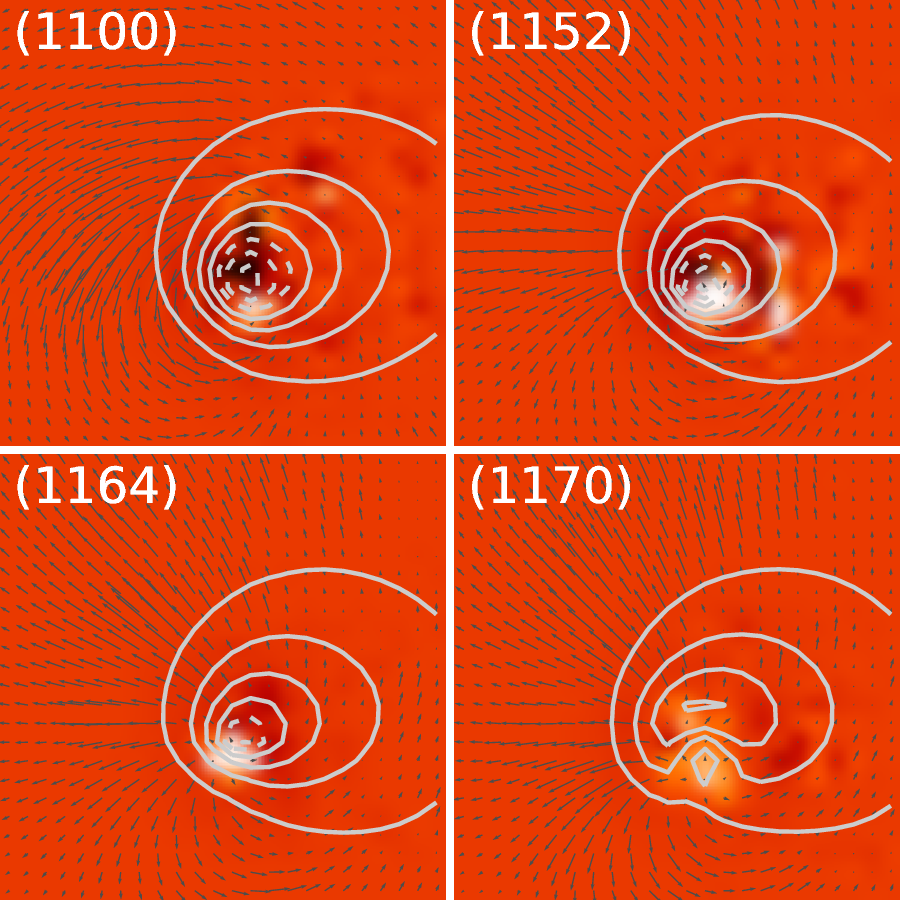}\\
\includegraphics[width=0.24\textwidth]{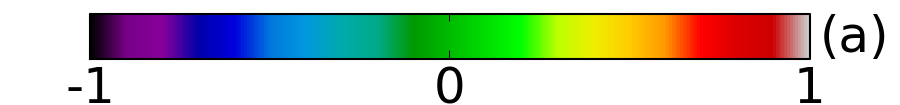}\hfill%
\includegraphics[width=0.24\textwidth]{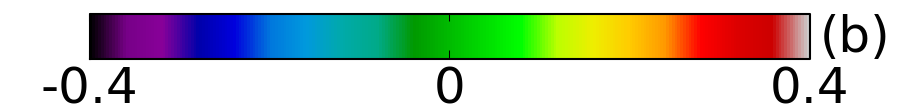}\hfill%
\includegraphics[width=0.24\textwidth]{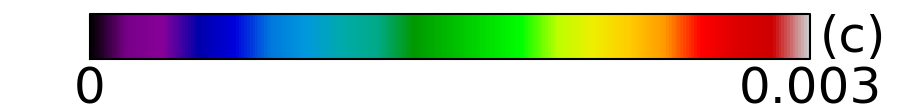}\hfill%
\includegraphics[width=0.24\textwidth]{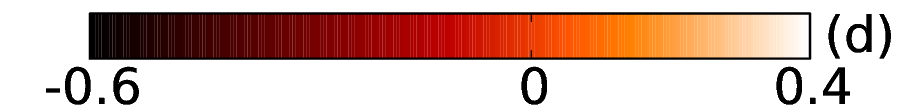}
\caption{\label{f:s1} Initial stage of the skyrmion annihilation process ($t=1152\,t_0$, see Fig.~\protect{\ref{f:Qt}}b): $\bm S$ (a), $\bm s$ (b), spin-torque $|\bm S\times \bm s|$ (c), and electron internal magnetic field $b$ Eq.~(\ref{eq:b}) at different times during the topological transition (d); $S_z$ contour lines (b-d); $(s_x,s_y)$ arrows (b-d); box-size is $64\times 64\,a^2$ for (a-c), and $24\times 24\,a^2$ for the (d) panel.}
\end{figure*}

We solved numerically the system (\ref{eq:LL}), (\ref{eq:c}) to compute the evolution of the magnetization texture $\bm S(\bm x,t)$ on the discrete lattice and its coupling with the itinerant spin density $\bm s(\bm x,t)$, and to monitor changes in the topological charge $Q(t)$. The Schrödinger equation is solved using an unitary time-splitting method, and the Landau-Lifshitz equation, a fourth order Runge-Kutta time stepping; difference operators are exactly computed on the lattice using a pseudo-spectral algorithm. The time step is tuned to reach machine precision conservation of the magnetization norm. The initial magnetization distribution is carefully determined to satisfy the stationary Landau-Lifshitz equation for vanishing electron current. In practice we start with an exact solution of the isotropic (continuous) magnetization equation ($K=D=J_s=0$), in the form of localized vortices, and let this distribution evolve in time with the full lattice Landau-Lifshitz equation (in zero external electric field $E=0$); the asymptotic stationary state is then used as the initial condition of the coupled system (\ref{eq:LL}), (\ref{eq:c}). We verified that eventual changes of an initial metastable state (triggered by numerical noise, for example) actually occur at times much longer than the ones observed after injection of the electron current; therefore, we may conclude that the observed changes (notably in the topology of the spin texture) result from the interaction with the spin polarized current. We present results of initial arrays of vortices and skyrmions, and of isolated structures like the Belavin-Polyakov skyrmion.

\begin{figure}[t]
\centering
\includegraphics[width=0.24\textwidth]{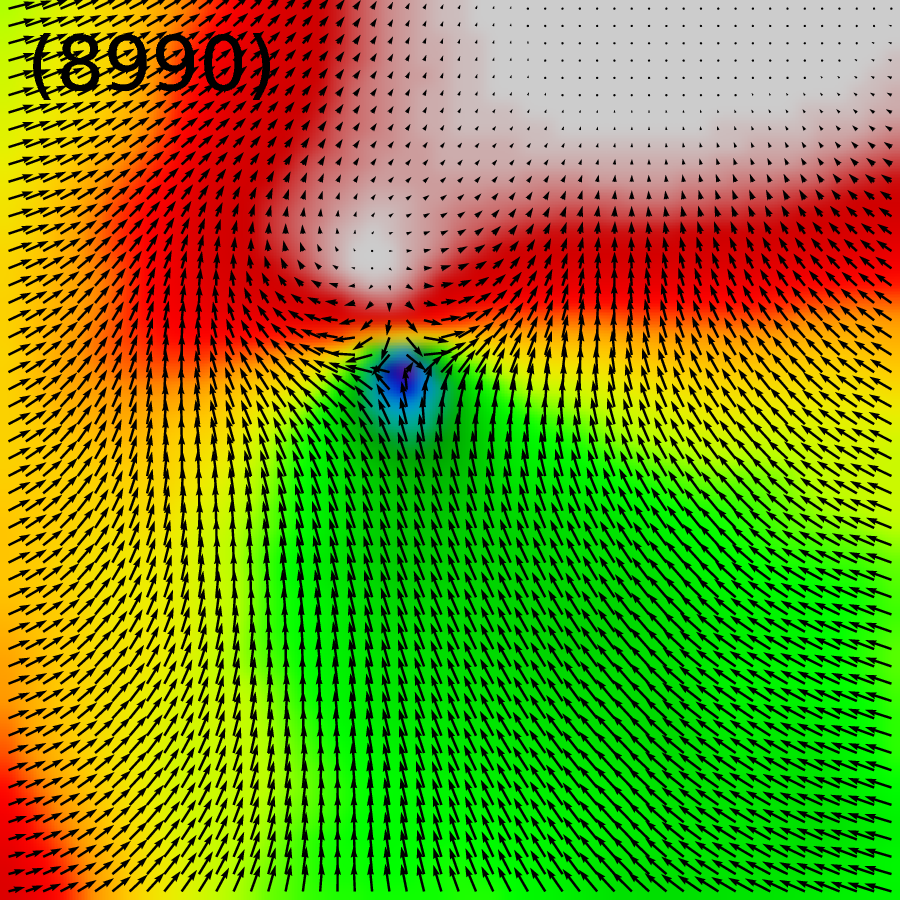}\hfill%
\includegraphics[width=0.24\textwidth]{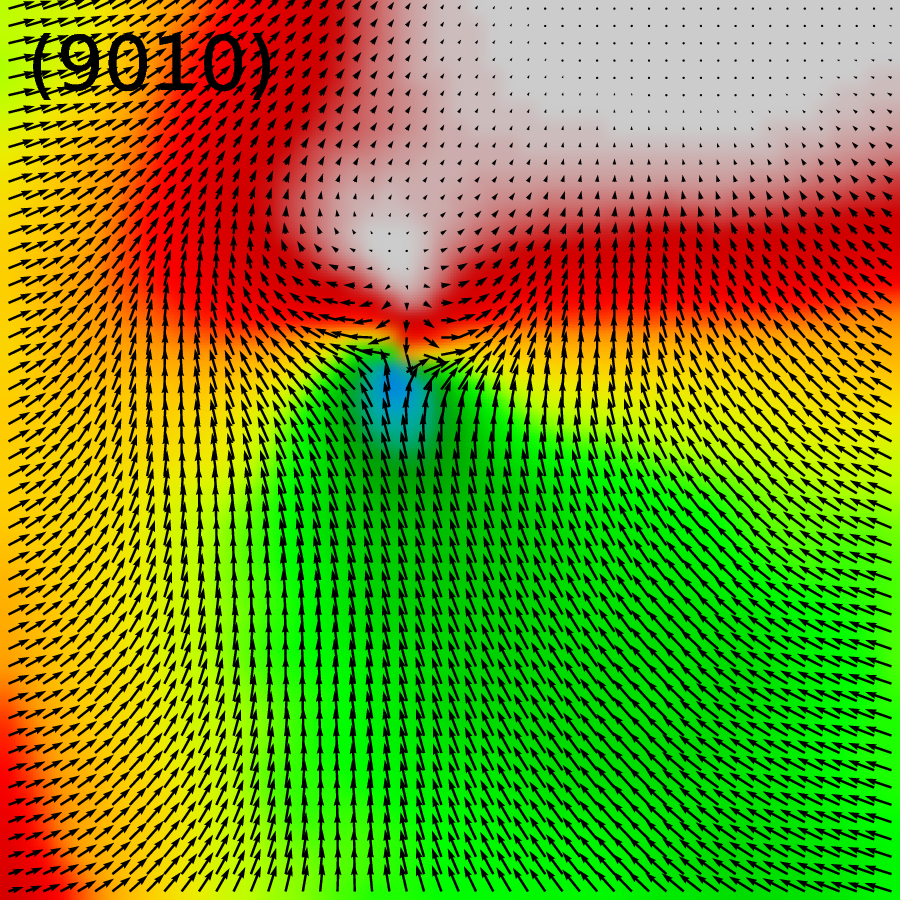}\\
\includegraphics[width=0.48\textwidth]{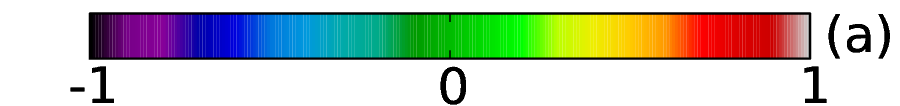}
\includegraphics[width=0.48\textwidth]{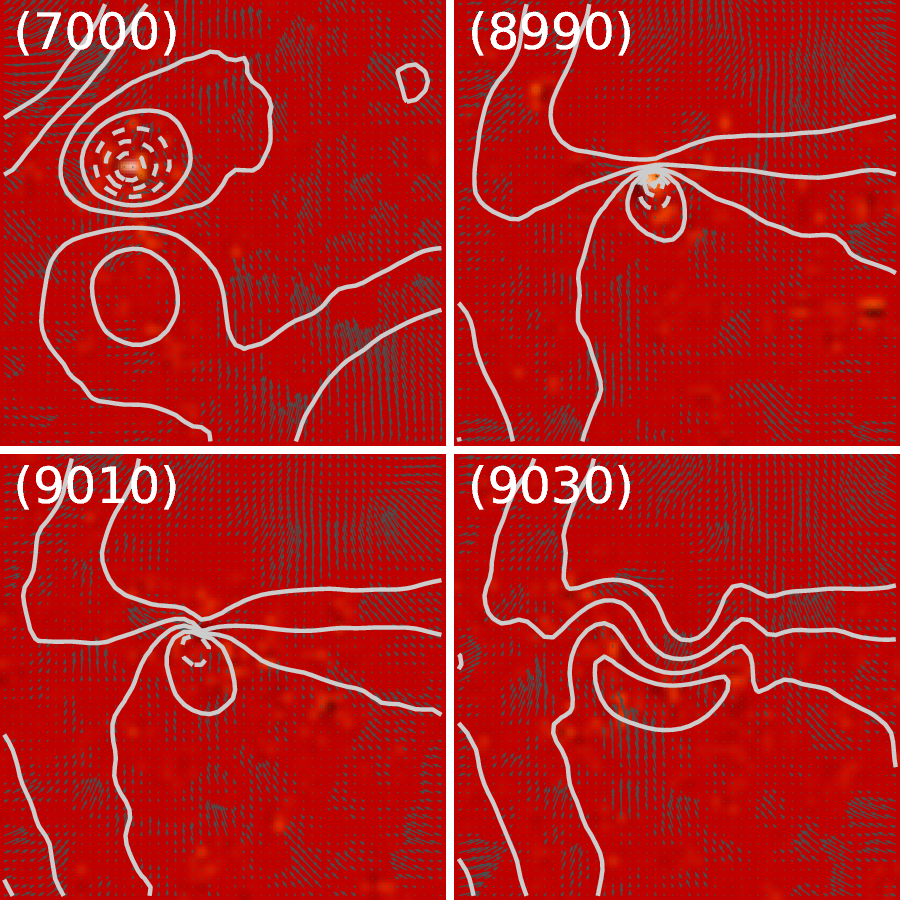}\\
\includegraphics[width=0.48\textwidth]{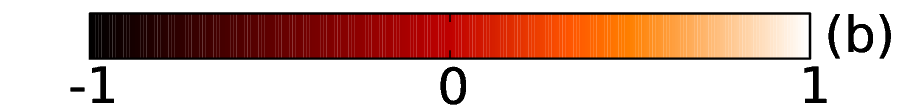}
\caption{\label{f:v0_b} (Color online) Magnetization and electron $b$-field for the vortex array case (Fig.~\protect\ref{f:S}). The annihilation of equal charge vortex-antivortex is accompanied by the nucleation of a strong electron vortex (black spot at $t=8990\,t_0$). The change $\Delta Q_+=4$ corresponds to the simultaneous destruction of four $Q=1/2$ vortex pairs. }
\end{figure}

\begin{figure}
\centering
\includegraphics[width=0.48\textwidth]{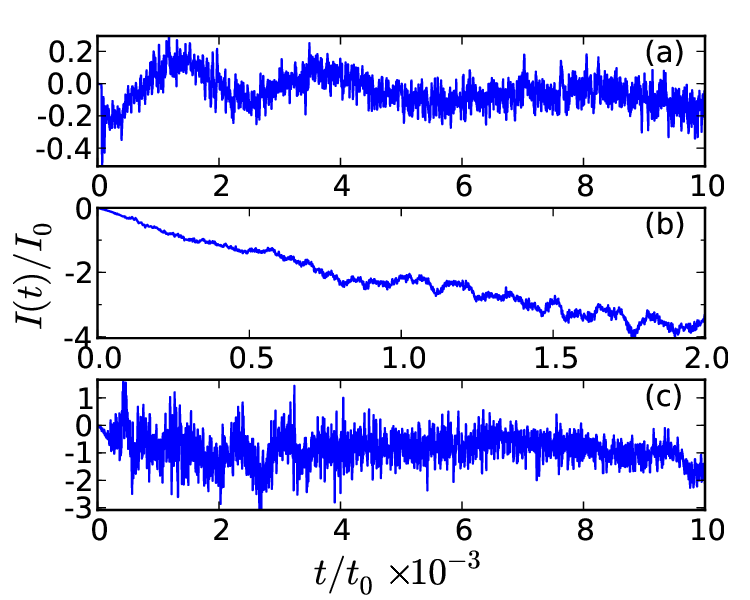}
\caption{\label{f:J} (Color online) Electron current for skirmion lattice (a), Bleavin-Polyakov (b) and vortex array (c) cases also shown in Fig.~\protect\ref{f:S}. ($I_0$, the unit of current, is in the range $10^{-4}$ to $10^{-5}\,\mathrm{A}$.)} 
\end{figure}

\begin{figure*}
\centering
\includegraphics[width=0.24\textwidth]{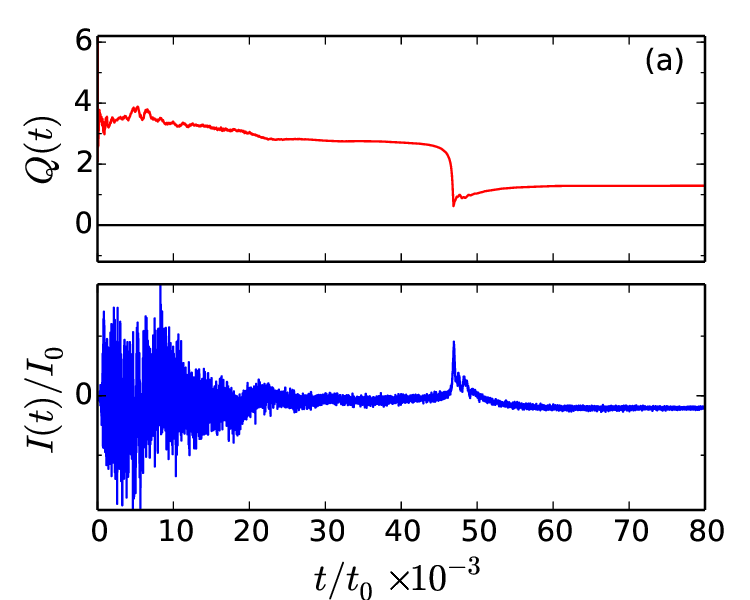}\hfill%
\includegraphics[width=0.24\textwidth]{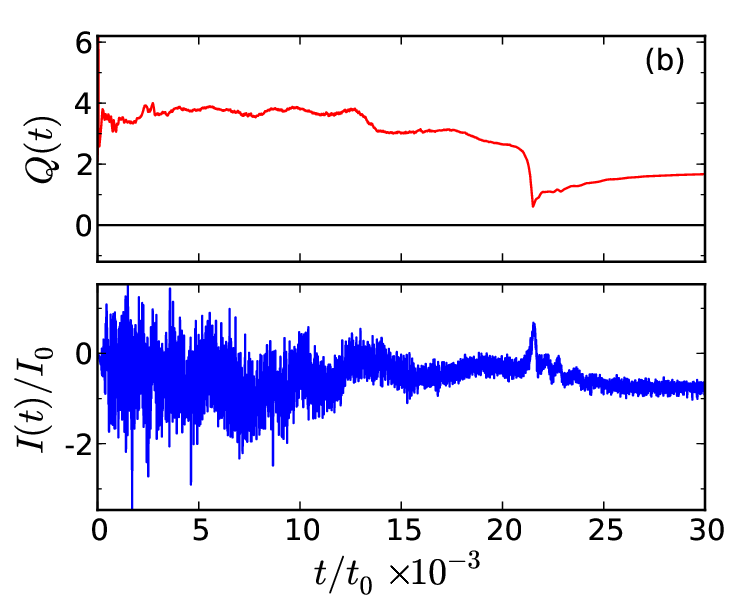}\hfill%
\includegraphics[width=0.24\textwidth]{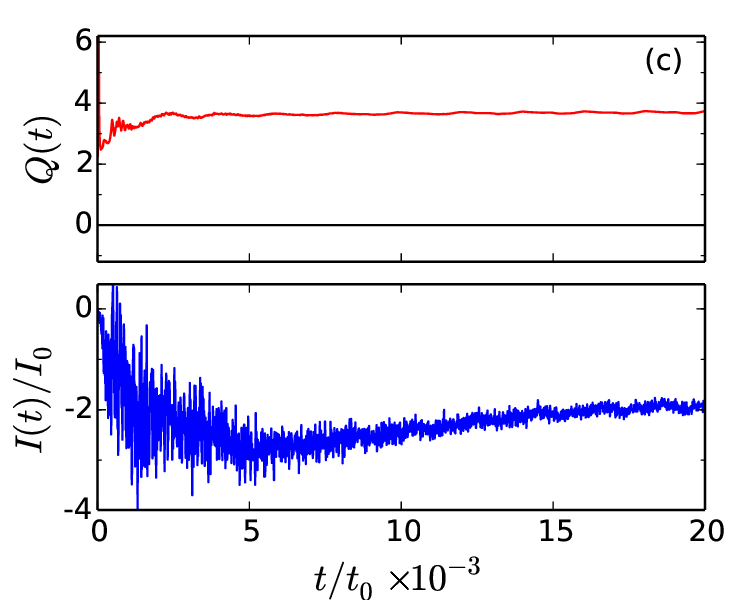}\hfill%
\includegraphics[width=0.24\textwidth]{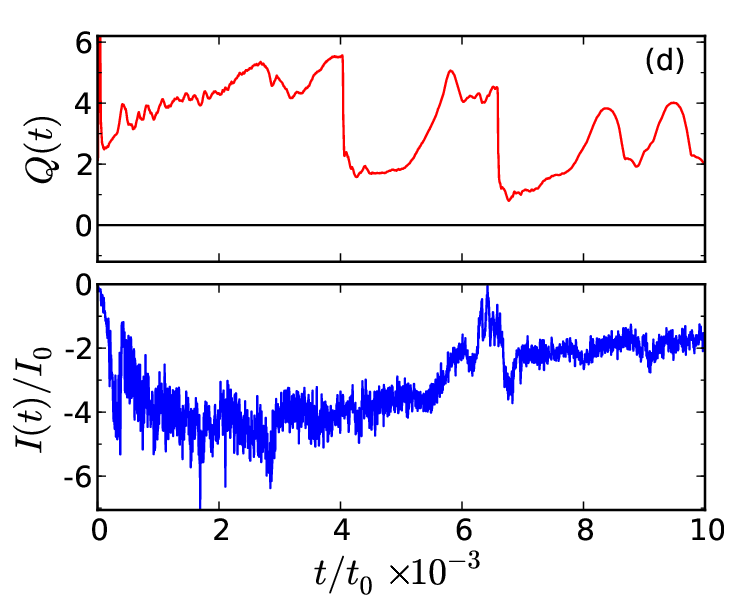}
\caption{\label{f:cj} (Color online) Topological charge (upper panel, in red $Q_+$, $Q=0$) and current (lower panel) for the vortex array case, as a function of the applied electric field ((a) $E=10^{-5}$, (b) $E=5\,10^{-5}$, (c) $E=2\,10^{-4}$, and (d) $E=5\,10^{-4}$, with $n_e=0.1$) and easy plane anisotropy ($K=0.01$ in (a) and (b), and $K=0.05$ in (c) and (d)). In (a-b), at variance to the case of Fig.~\ref{f:v0_b}, the change $\Delta Q_+=2$ corresponds to the trivial annihilation of two pairs of opposite charge vortex-antivortex.}
\end{figure*}

%
The typical parameters used in the simulations are $J_s=1$, $J=0.1,\,0.4$, $K=0,\,\pm 0.01$, $D=0,\,0.01$, for the coupling energies in units of $\epsilon$, and $\alpha=0.1$ for the Gilbert constant. Finite values of $D$ are relevant for the skyrmion lattice (Figs.~\ref{f:Qt}a and~\ref{f:S} top row); $K<0$ (easy-axis) is used for the Belavin-Polyakov skyrmion (Figs.~\ref{f:Qt}b and~\ref{f:s1}). For vortices arrays we have $K>0$ (easy-plane) and $D=0$ (Figs.~\ref{f:Qt}c,~\ref{f:S} bottom row, and~\ref{f:v0_b}). The electron density and electric field are chosen in order to get effective spin current densities of the order of the experimental ones ($10^{12}\,\mathrm{A\,m^{-2}}$ in three-dimensional ferromagnets)\cite{Yamada-2008eu}:  $n_e\approx 0.1$ electrons per site, $E/E_0\approx 10^{-5}\mathrm{-}5\,10^{-4}$, and a spin polarization $\mu_e B_p=0.1\,\epsilon$. We remark however, that in our model we neglect dissipation effects other than $\alpha$-relaxation, and as a consequence, the current density may increase over time (we fix the electric field); we will see below examples of its behavior (see Fig.~\ref{f:J} and Fig.~\ref{f:cj}).

As an illustration of the rich phenomenology exhibited by the Schrödinger-Landau-Lifshitz system, we show in Figs.~\ref{f:Qt} and~\ref{f:S} the topological evolution of skyrmions and vortices in the presence of initial free electrons subject to an external electric field.\footnote{See Supplemental Material at [URL] for movies showing the time evolution of the skyrmions and vortices driven by the spin-polarized current} The topological charge (\ref{eq:Q}) decreases or increases by integer steps $\Delta Q=1$. Variations with $\Delta Q>1$ result from the superposition of simultaneous and separated in space $\Delta Q=1$ events. In Fig.~\ref{f:Qt} we also plot $Q_+(t)$ computed from the integral of $|q|$, as in (\ref{eq:Q}), 
\[
Q_+=\int_{\mathbb R^2} \frac{d\bm x }{4\pi}|q(\bm x,t)|\,,
\]
which is a measure of the number of vortices present in the system at time $t$ (at variance with $Q$, it is not a conserved quantity of the Landau-Lifshitz dynamics). In these examples we used a strong electric field ($E/E_0 = 10^{-4}$, equivalent to $10^{5}\,\mathrm{V\,m^{-1}}$) in order to clearly display the current-vortex interactions.

%

%

We may distinguish two ways leading to a topological change, according to the value of $\Delta Q$: The nucleation and annihilation of same polarity vortex-antivortex pairs that do not change the total topological charge, $\Delta Q=0$; and the reversal of a vortex core, the suppression of a skyrmion, or other vortex interactions involving a change $\Delta Q=1$. Figure~\ref{f:S} shows the magnetization at selected times, for an initial skyrmion lattice and an array of vortices, displaying a variety of topological change events. 

The skyrmion lattice may be considered as a superposition of bounded meron-antimeron pairs,\cite{Gross-1978kx} double-periodically distributed in the plane, and having a charge $Q=8\times(1/2)$ per cell. Under the action of a strong $+x$-spin electron current, they wander around as almost independent $Q=1/2$ structures (Fig.~\ref{f:S}, $t=2000$), and when equal charge pairs come close together, they annihilate emitting a burst of spin waves (Fig.~\ref{f:S}, $t=4680,\,4760$ and $t=9004,\,9016$). Lately, a uniform magnetization state is reached (Fig.~\ref{f:S}, $t=9600$). The annihilation events are clearly identified by a discontinuity $\Delta Q=-2$, as can be seen in Fig.~\ref{f:Qt}a.

The vortex array is a superposition of vortex-antivortex opposite sign pairs, the total topological charge is then $Q=0$. Even if the total charge vanishes, the dynamics and interactions of individual structures are highly non trivial. Subject to a spin-up current, the vortex array evolution (Fig.~\ref{f:S}) in addition to the vortex annihilation event at times $t=8990$--$9030$, which is similar to the one observed in the skyrmion case, shows other interesting processes, such as the nucleation of vortex-antivortex $Q=0$ pairs ($t=4500$--$7000$) from magnetization structures created by the polarized electrons (white-red patches). Each of these events is easily correlated to a sudden change in $Q_+$, Fig.~\ref{f:Qt}c.

In general, the mechanism of a $\Delta Q=1$ topological change entails the formation of a virtual structure (a singularity in the continuum limit) with a net unit charge, opposite to the initial charge. For instance, in the case of a $Q=1/2$ vortex, a bump of opposite polarity forms near the core, even when the vortex is almost at rest, as a result of the torque exerted by the electrons. At the time when the two peaks approach close enough, at most a few lattice steps away, a virtual antivortex should appear to provide the necessary charge to annihilate the original vortex, then allowing the growing bump to become a new vortex. This pathway through the $\Delta Q=1$ change is, with respect to the topology of the magnetization field, similar to the core reversal phenomenology observed in micromagnetic simulations for moving\cite{Hertel-2007jy,Guslienko-2008so} or static structures.\cite{Kravchuk-2009vn,Kammerer-2011fk} However, at variance to these models where the driven mechanism is an external time-dependent magnetic field, here we take into account the self-consistent interaction with the electron current (maintained by an external, constant, electric field). In addition to the precession impressed around the local direction of the spin-polarized current, the action of the moving electrons $\bm s$ on the fixed spins texture is twofold: first they reduce the vortex core size, through a non-local interaction with the surrounding spin currents and waves; and second, they are able, by a local spin-transfer torque, to reverse the orientation of individual spins (strong non-adiabatic effect). The  annihilation of a skyrmion core, presented in Fig.~\ref{f:s1}, is significant of the role played by the itinerant spins in the $\Delta Q=1$ topological change.

%

Figure~\ref{f:s1} presents the configuration of the Belavin-Polyakov skyrmion in the initial stage of the topological change, corresponding to $t=1152$ in Fig.~\ref{f:Qt}a. The skyrmion core was previously deformed by the spin-up polarized current, increasing the gradient of $S_z$ in the $x$-direction (Fig.~\ref{f:s1}a); as in the case of a meron core switching, the core of the skyrmion is ultimately reversed, leaving a $Q=0$ final state (at time $t=1168$). One of the main characteristics of the free spins is its quasi-stochastic distribution, as one observes in Fig.~\ref{f:s1}b. Reminding that the differentiability of the effective field $\bm f$, is a necessary condition for the conservation of the topological charge, the spatio-temporal intermittency of the $\bm s=\langle c^\dag \bm \sigma c\rangle$ field is a crucial ingredient in the microscopic mechanism of the topological change. The origin of this complex behavior is the multiple quantum scattering of the electron waves on the magnetization inhomogeneities, as can be verified by following the evolution of their wave function (compare the fixed and itinerant spin distributions of Fig.~\ref{f:s1}). We also show in Fig.~\ref{f:s1} the effective internal magnetic field created by the fluctuating spin texture of the itinerant electrons,
\begin{equation}\label{eq:b}
b=\bm n\cdot\partial_x\bm n 
	\times \partial_y \bm n,\quad \bm n=\bm s/|\bm s|\,.
\end{equation}
It arises when imposing the electron spin direction as the natural quantization axis, leading to an effective gauge vector potential $\bm a=(\bm n\times \nabla \bm n) \cdot \bm \sigma$. This is opposite to the usual gauge transformation that takes the magnetization as the reference frame, to locally rotate the quantization axis (used to eliminate the electron degrees of freedom from the action).\cite{Tatara-2008kx,Zang-2011fk} The remarkable fact about this quantity is that it concentrates at the vortex core, presenting a strong gradient precisely in the region where the core is reversing. The relation between the internal $b$ field and the topological charge density of the electron spin field, allows us to interpret the rotating opposite polarity peaks at the center of the vortex core, as being the signature of a nontrivial topological structure that will trigger the formation of the $Q=1$ extra charge necessary to the transition. Therefore, this process is in some sense the opposite of the quasi-adiabatic mechanism: the magnetization vector, at a microscopic spatio-temporal scale, follows the dynamics imposed by the itinerant electron spins, which are the source of the topological change. The formation of an electron vortex, revealed by the presence of localized structures with a strong $b$-field, is the main result of this paper. This is further evidenced by the observation of the topological change under rather different conditions, as in the annihilation of a vortex-antivortex pair.

In the case of the vortex array, the $z$-spin polarized current drives the formation of vortex-antivortex pairs, which detach from the large polarized current aligned patches, as presented in Fig.~\ref{f:S}. (The current density is strongly inhomogeneous, even for very weak electric fields.) These vortices then interact with the original vortex array, and eventually annihilate with its equal sign pair. The detail of the magnetization field around the annihilation time ($t\approx 8990\,t_0$), and the corresponding $b$-electron field are presented in Fig.~\ref{f:v0_b}. The original negative polarity vortex (dashed contour lines at time 7000, Fig.~\ref{f:v0_b}b) and the nucleated positive polarity vortex (both having $Q=1/2$) turn around each other; the negative polarity one is associated with a positive $b$-field. At time 8990 the two vortices rotated of about $\pi$, and a spot of negative $b$-field appears. From the colormap we note that this electron vortex possesses a $-1$ charge, opposite to the $Q=1$ charge of the former vortex-antivortex pair. After the annihilation, a strong emission of spin waves is observed, and the $b$-field concentrations disappear.

From this rich phenomenology of topological changes we may distinguish (see Table~\ref{t:dq}): (i) the destruction of Belavin-Polyakov skyrmion subject to a spin current with polarization antiparallel to its core magnetization, (ii) the equal sign vortices annihilation and, (iii) the opposite sign vortices annihilation. In the (i) and (ii) cases, for which the topological charge change, in spite of their differences the $\Delta Q=-1$ driving microscopic mechanism is related to the formation of a localized electron spin structure (compare Fig.~\ref{f:s1} and Fig.~\ref{f:v0_b}). Indeed, the superposition of the equal sign vortex-antivortex of Fig.~\ref{f:v0_b} is topologically similar to the Belavin-Polyakov skyrmion, both having $Q=1$ (or $Q=-1$).

The electric current in units of $I_0$ is computed from the formula 
\begin{equation}\label{eq:I}
    I(t) = -\sum_{y}\bigg\langle 
        \frac{\partial H_e}{\partial A}\bigg\rangle
\end{equation}
where the mean is taken over the quantum state $c_i(t)$ for all $i=(x_0,y)$ lattice sites at a fixed $x=x_0$ position ($A = Et$ is the $x$ component of the vector potential). It is plotted in Fig.~\ref{f:J} for the three cases presented in Fig.~\ref{f:Qt}. We remark that in spite of the dissipationless dynamics of the electrons, the current is not simply linear in time, it may even be almost suppressed as in the skyrmion lattice case. In the case of the Belavin-Polyakov skyrmion, the current increases drastically only when the vortex core disappears and an almost uniform magnetization background remains. The actual evolution of the electric current drastically depends on the magnetization distribution: the three initial textures lead to very different current behavior and characteristic orders of magnitude. 

We also observed that the electronic density has a minimum inside the vortex cores, and tends to concentrate between the vortex structures, in regions with a smooth varying magnetization. From this observation one may think that under a certain threshold, the polarized current would be unable to drive the topological changes, because of its weak interaction with the vortices. To investigate this point, we performed a series of simulations of the vortex array for different values of the electric field and the anisotropy constant. Some representative results are shown in Fig.~\ref{f:cj} where we plot the topological charge and the electric current as a function of time. For weak easy-plane anisotropy ($K=0.01$, Fig.~\ref{f:cj}a and~\ref{f:cj}b) and weak electric field ($E \ll 10^{-4}$), the nucleation of vortices by the polarized current is suppressed; however, it is able to slowly drive the opposite sign original vortices close together, allowing their $\Delta Q = 0$ annihilation (times near 50000 for $E=10^{-5}$ and 20000 for $E=5\,10^{-5}$). The increase of the easy plane anisotropy ($K=0.05$, Fig.~\ref{f:cj}c and~\ref{f:cj}d) contributes to the stabilization of the vortex array, and stronger electric fields are needed to change the topology. A stationary state naturally arises in the case $K=0.05$ and $E=2\,10^{-4}$ of Fig.~\ref{f:cj}c. For a stronger electric field, $E=5\,10^{-4}$, a rich dynamics develops, with nucleation and annihilation of vortices (Fig.~\ref{f:cj}d).

%
\section{Discussion and conclusion}

\begin{table}
    \caption{\label{t:dq}Elementary topology transformations.}
    \begin{ruledtabular}
        \begin{tabular}{llccl}
            \#    & Configuration         & $\Delta Q$ & $\Delta Q_+$ & 
            Description                                   \\
(i)   & skyrmion annihilation & $1$        & $-1$         & 
            b-field structure                             \\
            (ii)  & V-AV\footnotemark[1]\footnotetext{%
    Vortex-antivortex} 
            annihilation      & $-1$       & $-1$         & 
            collision, b-field                            \\
(iii) & V-AV nucleation       & $0$        & $1$          & 
            current gradients                             \\
(iv)  & V-AV annihilation     & $0$        & $-1$         & 
            collision, smooth                             \\
        \end{tabular}
    \end{ruledtabular}
\end{table}

This paper focused on the influence of the itinerant electrons dynamics on the spin-transfer torque and the mechanisms of topological changes in two-dimensional magnetic textures. We proposed a simple model with a single coupling parameter between electrons and fixed magnetic moments, $J_{s}$. Other parameters take into account the exchange, anisotropy and spin-orbit effects. The unit of energy, which we took as $\epsilon\sim J_{s}$, and the unit of length $a$, can be considered as effective parameters whose values depend on the actual physical system. The $J_{s}$ coupling is typically of the order of the $1\,\mathrm{eV}$ in magnetic metals and one order of magnitude smaller in chiral magnets or diluted magnetic semiconductors; the exchange constant is about $J/J_{s}=0.1$.\cite{Berger-1984uq,Grigoriev-2006ys,Stohr-2006qf}

The elementary processes involving a topological change as found in our simulations, are summarized in Table \ref{t:dq}. Configuration (i) corresponds to the annihilation of a skyrmion core; as shown in Fig.~\ref{f:s1} this process involves an intermediate state characterized by an electron-spin structure that can be revealed by the topological $b$-field (\ref{eq:b}). A $b$-field structure is also present in the annihilation of a nontrivial configuration (ii), of a vortex-antivortex pair. Vortices of equal charge annihilate during collisions, and are followed by a burst of spin waves, as illustrated by the simulations of the skyrmion and vortex arrays (see Fig.~\ref{eq:LL} and in the supplemental material\cite{Note1} the corresponding movies). In contrast, topological trivial configurations evolve smoothly, like (iii) and (iv), corresponding respectively to nucleation and annihilation of vortex-antivortex pairs of opposite charges. Vortices are generated in pairs by traveling blobs of spin polarized electrons. Indeed, strong current inhomogeneities naturally appear by interaction of an initially homogeneous current with existent vortices; further evolution of these inhomogeneities leads to the nucleation of V-AV pairs. These elementary processes combine to display a complex dynamics; for example, in the case of vortex arrays, nucleation of pairs at the edges of electron blobs is followed by their interaction with preexisting vortices; or isolated current inhomogeneities in an essentially uniform background, create a cascade of trivial and nontrivial pairs of vortices that subsequently annihilate (see supplemental material,\cite{Note1} the movie corresponding to the vortex array of Fig.~\ref{f:cj}d).

If one compares these results with previous micromagnetic simulations including the spin torque term,\cite{Liu-2007hw,Nakatani-2008pz,Tchoe-2012uq} one finds that the phenomenology observed in the self-consistent case is much richer, even if micromagnetism can capture some of the elementary mechanisms of nucleation and annihilation. In order to drive topological changes in the micromagnetic framework, a current pulse\cite{Liu-2007hw} or an imposed stationary current inhomogeneity,\cite{Tchoe-2012uq} were used. The injection of a current pulse in a ferromagnet, induces the switching of a vortex core, through the nucleation of a vortex-antivortex pair subsequently annihilated by the formation of a Bloch point.\cite{Liu-2007hw} This process is similar to the case (ii), where the electron spin texture plays the role of intermediate structure (a Bloch point is intrinsically three-dimensional, and then it cannot be realized in our system; however a virtual superposition of a vortex and an antivortex having the same charge, is topologically equivalent to a Bloch point). Nucleation events as in case (iii) were also observed when the system is driven by a rotating inhomogeneous current density.\cite{Tchoe-2012uq} In this case, the nontrivial topology of the current density is transferred by spin-torque to the magnetization texture. In our simulations we find that the current-magnetic texture interaction is accompanied by a strong modification of the current distribution itself, leading sometimes to quasi-stationary states (as in the cases shown in Fig.~\ref{f:cj}a-c). Consideration of the electron dynamics allows to show first that the hypothesis of uniform current breaks down in the vicinity of vortex cores; that quantum transmission effects limit the current flow even without explicit dissipation effects; and finally, that strongly non-adiabatic processes (notably through the electron $b$-field) are essential for the occurrence of topological changes of skyrmions and equal charge vortex-antivortex pairs.

The scattering of electrons off magnetization inhomogeneities, notably vortex cores, is at the origin of the current spatial variations and the formation of polarized electron blobs. We observed that vortex arrays are stabilized by easy axis anisotropy, leading to a state characterized by a quasi-periodic emission of vortices (as in Fig.~\ref{f:cj}d). One of the reasons of this stabilization is that the propagation of $z$-polarized electrons is limited in an easy-plane medium: stronger the anisotropy, stronger the applied electric field needed for generating the same current. Decreasing the electric field, topological changes are suppressed, and an inhomogeneous stationary state settles in. The electric field threshold for the vortex array is found in the interval $E/E_0=(2,5)\,10^{-4}$ (compare Fig.~\ref{f:cj}c and Fig.~\ref{f:cj}d). However, the actual threshold values depend strongly on the anisotropy and other material and geometric parameters. In the case of the Belavin-Polyakov skyrmion our simulations show stability below electric fields of the order of $E/E_0 \approx 10^{-5}$, in the isotropic case $K=0$. Qualitatively we note that topological changes produce for currents $I \gtrsim 2 I_0$. These large currents are present in the case of the annihilation of the Belavin-Polyakov skyrmion, or in the nucleation of vortices by current blobs. In fact, lower currents can also induce topological changes, but indirectly, because of their effect on the motion of vortices that eventually collide, as in the case of the skyrmion lattice (note that the currents are of the order $I\approx 0.1 I_0$). Suppression of nucleation of new vortices from current inhomogeneities appears to occur below $I\approx 2 I_0$. Using a typical value for the current unit in ferromagnet, $I_0\approx 5\,10^{-5}\,\mathrm{A}$, and a rough estimation for the current density as $j\sim I/(La)$, one obtains an order of magnitude $j\sim 10^{12}\,\mathrm{Am^{-2}}$. At variance, in the case of a skyrmion lattice we did not found a minimum current below which vortices remain static; even currents of the order of $I/I_0\sim 10^{-2}$ (with $E/E_0= 10^{-5}$) are able to induce a skyrmion motion.

In summary, we investigated the topological changes in a two-dimensional ferromagnet driven by a self-consistent electron current. The Landau-Lifshitz equation is coupled through the spin-transfer torque term with the Schrödinger equation for the itinerant spins. At variance to the continuous micromagnetic models, the system discreteness and more importantly, the stochastic behavior of the driven term, broke the conservation of the topological charge. We observed that both, local and nonlocal interactions, play a role in the transition between different topological configurations. In particular, the electron current tends to concentrate in channels that avoid the vortex cores: strong gradients of the magnetization act as potential barriers, scattering off the electron waves. The phenomenology of a $\Delta Q=1$ change of an initial $Q=\pm 1,\,\pm 1/2$ vortex, although rich, reduces to a single topological mechanism, the nucleation of a $Q=\pm 1$ charge that annihilates the old structure, letting the new structure with the opposite charge or $Q=0$. The interesting point is that this mechanism do not arise spontaneously but is triggered, above a threshold, by the spin-polarized current. The electron spin and its associated polarized current are strongly fluctuating, up to the lattice and time unit scales, which appear to be the relevant scales for the topological changes in the dissipationless limit. The spatial inhomogeneity and localization of the electrons is a general feature, systematically observed, showing that the systems is relatively far from the quasi-adiabatic regime. The spontaneous nucleation and annihilation of vortices are in fact driven by the strong inhomogeneity of the electron spin distribution. 

At the heart of the topological change is the formation of an electron nontrivial structure that induces the switching mechanism of the magnetization in a strongly non-adiabatic process. This electron structure has a nontrivial topology characterized by an internal magnetic field. Localized spots of this field appear during the transition and are observed in apparently different processes such as the annihilation of the Belavin-Polyakov skyrmion or the interaction of equal sign vortices.

Our model is limited to a simple two dimensional geometry and boundary conditions; although it would be interesting to explore more complex situations, the main open question is how to obtain a continuous limit (necessary to model larger systems at longer time scales) that takes into account the effective dissipative coupling with the electrons (see the recent papers\cite{Zang-2011fk,Kim-2010uq} where dissipation and dissipationless mechanisms are analyzed). In spite of these limitations, the observation of topological changes driven by spin currents should be accessible to experiments with chiral magnets like MnSi, where the effectiveness of spin-transfer torque on skyrmion lattices was already demonstrated.\cite{Jonietz-2010ly} More recently,\cite{Romming-2013lq} it was proved that individual skyrmions can be created and annihilated by means of an electron current injected by a local probe (using the tip of a spin-polarized scanning tunneling microscope). The system undergoes a transition from a skyrmion to a ferromagnetic state, as the one shown in Fig.~\ref{f:Qt}b. The mechanism of this switching is attributed to a combination of nonthermal excitations of the injected electrons and the spin-transfer torque. Although a detailed comparison would require further investigation, these experimental results are in qualitative agreement with the scenario presented in Fig.~\ref{f:s1}. 

\begin{acknowledgments}
We thank Riccardo Hertel and Jean-Christophe Toussaint for useful discussions.
\end{acknowledgments}


\begin{thebibliography}{50}%
\makeatletter
\providecommand \@ifxundefined [1]{%
 \@ifx{#1\undefined}
}%
\providecommand \@ifnum [1]{%
 \ifnum #1\expandafter \@firstoftwo
 \else \expandafter \@secondoftwo
 \fi
}%
\providecommand \@ifx [1]{%
 \ifx #1\expandafter \@firstoftwo
 \else \expandafter \@secondoftwo
 \fi
}%
\providecommand \natexlab [1]{#1}%
\providecommand \enquote  [1]{``#1''}%
\providecommand \bibnamefont  [1]{#1}%
\providecommand \bibfnamefont [1]{#1}%
\providecommand \citenamefont [1]{#1}%
\providecommand \href@noop [0]{\@secondoftwo}%
\providecommand \href [0]{\begingroup \@sanitize@url \@href}%
\providecommand \@href[1]{\@@startlink{#1}\@@href}%
\providecommand \@@href[1]{\endgroup#1\@@endlink}%
\providecommand \@sanitize@url [0]{\catcode `\\12\catcode `\$12\catcode
  `\&12\catcode `\#12\catcode `\^12\catcode `\_12\catcode `\%12\relax}%
\providecommand \@@startlink[1]{}%
\providecommand \@@endlink[0]{}%
\providecommand \url  [0]{\begingroup\@sanitize@url \@url }%
\providecommand \@url [1]{\endgroup\@href {#1}{\urlprefix }}%
\providecommand \urlprefix  [0]{URL }%
\providecommand \Eprint [0]{\href }%
\providecommand \doibase [0]{http://dx.doi.org/}%
\providecommand \selectlanguage [0]{\@gobble}%
\providecommand \bibinfo  [0]{\@secondoftwo}%
\providecommand \bibfield  [0]{\@secondoftwo}%
\providecommand \translation [1]{[#1]}%
\providecommand \BibitemOpen [0]{}%
\providecommand \bibitemStop [0]{}%
\providecommand \bibitemNoStop [0]{.\EOS\space}%
\providecommand \EOS [0]{\spacefactor3000\relax}%
\providecommand \BibitemShut  [1]{\csname bibitem#1\endcsname}%
\let\auto@bib@innerbib\@empty
\bibitem [{\citenamefont {Cowburn}\ \emph {et~al.}(1999)\citenamefont
  {Cowburn}, \citenamefont {Koltsov}, \citenamefont {Adeyeye}, \citenamefont
  {Welland},\ and\ \citenamefont {Tricker}}]{Cowburn-1999vn}%
  \BibitemOpen
  \bibfield  {author} {\bibinfo {author} {\bibfnamefont {R.~P.}\ \bibnamefont
  {Cowburn}}, \bibinfo {author} {\bibfnamefont {D.~K.}\ \bibnamefont
  {Koltsov}}, \bibinfo {author} {\bibfnamefont {A.~O.}\ \bibnamefont
  {Adeyeye}}, \bibinfo {author} {\bibfnamefont {M.~E.}\ \bibnamefont
  {Welland}}, \ and\ \bibinfo {author} {\bibfnamefont {D.~M.}\ \bibnamefont
  {Tricker}},\ }\href {http://link.aps.org/abstract/PRL/v83/p1042} {\bibfield
  {journal} {\bibinfo  {journal} {Phys. Rev. Lett.}\ }\textbf {\bibinfo
  {volume} {83}},\ \bibinfo {pages} {1042} (\bibinfo {year}
  {1999})}\BibitemShut {NoStop}%
\bibitem [{\citenamefont {Shinjo}\ \emph {et~al.}(2000)\citenamefont {Shinjo},
  \citenamefont {Okuno}, \citenamefont {Hassdorf}, \citenamefont {Shigeto},\
  and\ \citenamefont {Ono}}]{Shinjo-2000dz}%
  \BibitemOpen
  \bibfield  {author} {\bibinfo {author} {\bibfnamefont {T.}~\bibnamefont
  {Shinjo}}, \bibinfo {author} {\bibfnamefont {T.}~\bibnamefont {Okuno}},
  \bibinfo {author} {\bibfnamefont {R.}~\bibnamefont {Hassdorf}}, \bibinfo
  {author} {\bibfnamefont {K.}~\bibnamefont {Shigeto}}, \ and\ \bibinfo
  {author} {\bibfnamefont {T.}~\bibnamefont {Ono}},\ }\href
  {http://www.sciencemag.org/cgi/content/abstract/289/5481/930} {\bibfield
  {journal} {\bibinfo  {journal} {Science}\ }\textbf {\bibinfo {volume}
  {289}},\ \bibinfo {pages} {930} (\bibinfo {year} {2000})}\BibitemShut
  {NoStop}%
\bibitem [{\citenamefont {Wachowiak}\ \emph {et~al.}(2002)\citenamefont
  {Wachowiak}, \citenamefont {Wiebe}, \citenamefont {Bode}, \citenamefont
  {Pietzsch}, \citenamefont {Morgenstern},\ and\ \citenamefont
  {Wiesendanger}}]{Wachowiak-2002pl}%
  \BibitemOpen
  \bibfield  {author} {\bibinfo {author} {\bibfnamefont {A.}~\bibnamefont
  {Wachowiak}}, \bibinfo {author} {\bibfnamefont {J.}~\bibnamefont {Wiebe}},
  \bibinfo {author} {\bibfnamefont {M.}~\bibnamefont {Bode}}, \bibinfo {author}
  {\bibfnamefont {O.}~\bibnamefont {Pietzsch}}, \bibinfo {author}
  {\bibfnamefont {M.}~\bibnamefont {Morgenstern}}, \ and\ \bibinfo {author}
  {\bibfnamefont {R.}~\bibnamefont {Wiesendanger}},\ }\href
  {http://www.sciencemag.org/cgi/content/abstract/298/5593/577} {\bibfield
  {journal} {\bibinfo  {journal} {Science}\ }\textbf {\bibinfo {volume}
  {298}},\ \bibinfo {pages} {577} (\bibinfo {year} {2002})}\BibitemShut
  {NoStop}%
\bibitem [{\citenamefont {Muhlbauer}\ \emph {et~al.}(2009)\citenamefont
  {Muhlbauer}, \citenamefont {Binz}, \citenamefont {Jonietz}, \citenamefont
  {Pfleiderer}, \citenamefont {Rosch}, \citenamefont {Neubauer}, \citenamefont
  {Georgii},\ and\ \citenamefont {Boni}}]{Muhlbauer-2009vn}%
  \BibitemOpen
  \bibfield  {author} {\bibinfo {author} {\bibfnamefont {S.}~\bibnamefont
  {Muhlbauer}}, \bibinfo {author} {\bibfnamefont {B.}~\bibnamefont {Binz}},
  \bibinfo {author} {\bibfnamefont {F.}~\bibnamefont {Jonietz}}, \bibinfo
  {author} {\bibfnamefont {C.}~\bibnamefont {Pfleiderer}}, \bibinfo {author}
  {\bibfnamefont {A.}~\bibnamefont {Rosch}}, \bibinfo {author} {\bibfnamefont
  {A.}~\bibnamefont {Neubauer}}, \bibinfo {author} {\bibfnamefont
  {R.}~\bibnamefont {Georgii}}, \ and\ \bibinfo {author} {\bibfnamefont
  {P.}~\bibnamefont {Boni}},\ }\href {\doibase 10.1126/science.1166767}
  {\bibfield  {journal} {\bibinfo  {journal} {Science}\ }\textbf {\bibinfo
  {volume} {323}},\ \bibinfo {pages} {915} (\bibinfo {year}
  {2009})}\BibitemShut {NoStop}%
\bibitem [{\citenamefont {Yu}\ \emph {et~al.}(2010)\citenamefont {Yu},
  \citenamefont {Onose}, \citenamefont {Kanazawa}, \citenamefont {Park},
  \citenamefont {Han}, \citenamefont {Matsui}, \citenamefont {Nagaosa},\ and\
  \citenamefont {Tokura}}]{Yu-2010fk}%
  \BibitemOpen
  \bibfield  {author} {\bibinfo {author} {\bibfnamefont {X.~Z.}\ \bibnamefont
  {Yu}}, \bibinfo {author} {\bibfnamefont {Y.}~\bibnamefont {Onose}}, \bibinfo
  {author} {\bibfnamefont {N.}~\bibnamefont {Kanazawa}}, \bibinfo {author}
  {\bibfnamefont {J.~H.}\ \bibnamefont {Park}}, \bibinfo {author}
  {\bibfnamefont {J.~H.}\ \bibnamefont {Han}}, \bibinfo {author} {\bibfnamefont
  {Y.}~\bibnamefont {Matsui}}, \bibinfo {author} {\bibfnamefont
  {N.}~\bibnamefont {Nagaosa}}, \ and\ \bibinfo {author} {\bibfnamefont
  {Y.}~\bibnamefont {Tokura}},\ }\href {http://dx.doi.org/10.1038/nature09124}
  {\bibfield  {journal} {\bibinfo  {journal} {Nature}\ }\textbf {\bibinfo
  {volume} {465}},\ \bibinfo {pages} {901} (\bibinfo {year}
  {2010})}\BibitemShut {NoStop}%
\bibitem [{\citenamefont {Heinze}\ \emph {et~al.}(2011)\citenamefont {Heinze},
  \citenamefont {von Bergmann}, \citenamefont {Menzel}, \citenamefont {Brede},
  \citenamefont {Kubetzka}, \citenamefont {Wiesendanger}, \citenamefont
  {Bihlmayer},\ and\ \citenamefont {Blugel}}]{Heinze-2011fk}%
  \BibitemOpen
  \bibfield  {author} {\bibinfo {author} {\bibfnamefont {S.}~\bibnamefont
  {Heinze}}, \bibinfo {author} {\bibfnamefont {K.}~\bibnamefont {von
  Bergmann}}, \bibinfo {author} {\bibfnamefont {M.}~\bibnamefont {Menzel}},
  \bibinfo {author} {\bibfnamefont {J.}~\bibnamefont {Brede}}, \bibinfo
  {author} {\bibfnamefont {A.}~\bibnamefont {Kubetzka}}, \bibinfo {author}
  {\bibfnamefont {R.}~\bibnamefont {Wiesendanger}}, \bibinfo {author}
  {\bibfnamefont {G.}~\bibnamefont {Bihlmayer}}, \ and\ \bibinfo {author}
  {\bibfnamefont {S.}~\bibnamefont {Blugel}},\ }\href
  {http://dx.doi.org/10.1038/nphys2045} {\bibfield  {journal} {\bibinfo
  {journal} {Nature Phys.}\ }\textbf {\bibinfo {volume} {7}},\ \bibinfo {pages}
  {713} (\bibinfo {year} {2011})}\BibitemShut {NoStop}%
\bibitem [{\citenamefont {Belavin}\ and\ \citenamefont
  {Polyakov}(1975)}]{Belavin-1975xw}%
  \BibitemOpen
  \bibfield  {author} {\bibinfo {author} {\bibfnamefont {A.~A.}\ \bibnamefont
  {Belavin}}\ and\ \bibinfo {author} {\bibfnamefont {A.~M.}\ \bibnamefont
  {Polyakov}},\ }\href@noop {} {\bibfield  {journal} {\bibinfo  {journal} {JETP
  Lett.}\ }\textbf {\bibinfo {volume} {22}},\ \bibinfo {pages} {245} (\bibinfo
  {year} {1975})}\BibitemShut {NoStop}%
\bibitem [{\citenamefont {Bogdanov}\ and\ \citenamefont
  {Yablonskii}(1989)}]{Bogdanov-1989fk}%
  \BibitemOpen
  \bibfield  {author} {\bibinfo {author} {\bibfnamefont {A.~N.}\ \bibnamefont
  {Bogdanov}}\ and\ \bibinfo {author} {\bibfnamefont {D.~A.}\ \bibnamefont
  {Yablonskii}},\ }\href@noop {} {\bibfield  {journal} {\bibinfo  {journal}
  {Sov. Phys. JETP}\ }\textbf {\bibinfo {volume} {68}},\ \bibinfo {pages} {101}
  (\bibinfo {year} {1989})},\ \bibinfo {note} {russian original - ZhETF, Vol.
  95, p. 178, 1989}\BibitemShut {NoStop}%
\bibitem [{\citenamefont {Dubrovin}\ \emph {et~al.}(1990)\citenamefont
  {Dubrovin}, \citenamefont {Fomenko},\ and\ \citenamefont
  {Novikov}}]{Dubrovin-1985fk}%
  \BibitemOpen
  \bibfield  {author} {\bibinfo {author} {\bibfnamefont {B.}~\bibnamefont
  {Dubrovin}}, \bibinfo {author} {\bibfnamefont {A.}~\bibnamefont {Fomenko}}, \
  and\ \bibinfo {author} {\bibfnamefont {S.}~\bibnamefont {Novikov}},\
  }\href@noop {} {\emph {\bibinfo {title} {Modern Geometry--methods and
  Applications: The geometry and topology of manifolds}}},\ Vol.~\bibinfo
  {volume} {2}\ (\bibinfo  {publisher} {Springer},\ \bibinfo {address}
  {Berlin},\ \bibinfo {year} {1990})\BibitemShut {NoStop}%
\bibitem [{\citenamefont {Kosevich}\ \emph {et~al.}(1990)\citenamefont
  {Kosevich}, \citenamefont {Ivanov},\ and\ \citenamefont
  {Kovalev}}]{Kosevich-1990cy}%
  \BibitemOpen
  \bibfield  {author} {\bibinfo {author} {\bibfnamefont {A.~M.}\ \bibnamefont
  {Kosevich}}, \bibinfo {author} {\bibfnamefont {B.~A.}\ \bibnamefont
  {Ivanov}}, \ and\ \bibinfo {author} {\bibfnamefont {A.~S.}\ \bibnamefont
  {Kovalev}},\ }\href
  {http://www.sciencedirect.com/science/article/B6TVP-46SXP0P-19/2/d0a99bbbccf078602123db3e5d601202}
  {\bibfield  {journal} {\bibinfo  {journal} {Phys. Rep.}\ }\textbf {\bibinfo
  {volume} {194}},\ \bibinfo {pages} {117} (\bibinfo {year}
  {1990})}\BibitemShut {NoStop}%
\bibitem [{\citenamefont {Tretiakov}\ and\ \citenamefont
  {Tchernyshyov}(2007)}]{Tretiakov-2007fc}%
  \BibitemOpen
  \bibfield  {author} {\bibinfo {author} {\bibfnamefont {O.~A.}\ \bibnamefont
  {Tretiakov}}\ and\ \bibinfo {author} {\bibfnamefont {O.}~\bibnamefont
  {Tchernyshyov}},\ }\href {http://link.aps.org/abstract/PRB/v75/e012408}
  {\bibfield  {journal} {\bibinfo  {journal} {Phys. Rev. B}\ }\textbf {\bibinfo
  {volume} {75}},\ \bibinfo {pages} {012408} (\bibinfo {year}
  {2007})}\BibitemShut {NoStop}%
\bibitem [{\citenamefont {Gross}(1978)}]{Gross-1978kx}%
  \BibitemOpen
  \bibfield  {author} {\bibinfo {author} {\bibfnamefont {D.~J.}\ \bibnamefont
  {Gross}},\ }\href {\doibase http://dx.doi.org/10.1016/0550-3213(78)90470-4}
  {\bibfield  {journal} {\bibinfo  {journal} {Nuclear Physics B}\ }\textbf
  {\bibinfo {volume} {132}},\ \bibinfo {pages} {439} (\bibinfo {year}
  {1978})}\BibitemShut {NoStop}%
\bibitem [{\citenamefont {Senthil}\ \emph {et~al.}(2004)\citenamefont
  {Senthil}, \citenamefont {Vishwanath}, \citenamefont {Balents}, \citenamefont
  {Sachdev},\ and\ \citenamefont {Fisher}}]{Senthil-2004kx}%
  \BibitemOpen
  \bibfield  {author} {\bibinfo {author} {\bibfnamefont {T.}~\bibnamefont
  {Senthil}}, \bibinfo {author} {\bibfnamefont {A.}~\bibnamefont {Vishwanath}},
  \bibinfo {author} {\bibfnamefont {L.}~\bibnamefont {Balents}}, \bibinfo
  {author} {\bibfnamefont {S.}~\bibnamefont {Sachdev}}, \ and\ \bibinfo
  {author} {\bibfnamefont {M.~P.~A.}\ \bibnamefont {Fisher}},\ }\href {\doibase
  10.1126/science.1091806} {\bibfield  {journal} {\bibinfo  {journal}
  {Science}\ }\textbf {\bibinfo {volume} {303}},\ \bibinfo {pages} {1490}
  (\bibinfo {year} {2004})}\BibitemShut {NoStop}%
\bibitem [{\citenamefont {Usov}\ and\ \citenamefont
  {Peschany}(1993)}]{Usov-1993oq}%
  \BibitemOpen
  \bibfield  {author} {\bibinfo {author} {\bibfnamefont {N.~A.}\ \bibnamefont
  {Usov}}\ and\ \bibinfo {author} {\bibfnamefont {S.~E.}\ \bibnamefont
  {Peschany}},\ }\href
  {http://www.sciencedirect.com/science/article/B6TJJ-46FHH1R-C8/1/8ec0a8b6c6db7ecfc8283c4b6fee09e1}
  {\bibfield  {journal} {\bibinfo  {journal} {J. Magn. Magn. Mater.}\ }\textbf
  {\bibinfo {volume} {118}},\ \bibinfo {pages} {L290} (\bibinfo {year}
  {1993})}\BibitemShut {NoStop}%
\bibitem [{\citenamefont {Berger}(1996)}]{Berger-1996bh}%
  \BibitemOpen
  \bibfield  {author} {\bibinfo {author} {\bibfnamefont {L.}~\bibnamefont
  {Berger}},\ }\href {http://link.aps.org/abstract/PRB/v54/p9353} {\bibfield
  {journal} {\bibinfo  {journal} {Phys. Rev. B}\ }\textbf {\bibinfo {volume}
  {54}},\ \bibinfo {pages} {9353} (\bibinfo {year} {1996})}\BibitemShut
  {NoStop}%
\bibitem [{\citenamefont {Slonczewski}(1996)}]{Slonczewski-1996lq}%
  \BibitemOpen
  \bibfield  {author} {\bibinfo {author} {\bibfnamefont {J.~C.}\ \bibnamefont
  {Slonczewski}},\ }\href
  {http://www.sciencedirect.com/science/article/B6TJJ-403WKFH-1/2/fd06569c67005bde75d947ee6cfc273e}
  {\bibfield  {journal} {\bibinfo  {journal} {J. Magn. Magn. Mater.}\ }\textbf
  {\bibinfo {volume} {159}},\ \bibinfo {pages} {L1} (\bibinfo {year}
  {1996})}\BibitemShut {NoStop}%
\bibitem [{\citenamefont {Van~Waeyenberge}\ \emph {et~al.}(2006)\citenamefont
  {Van~Waeyenberge}, \citenamefont {Puzic}, \citenamefont {Stoll},
  \citenamefont {Chou}, \citenamefont {Tyliszczak}, \citenamefont {Hertel},
  \citenamefont {Fahnle}, \citenamefont {Bruckl}, \citenamefont {Rott},
  \citenamefont {Reiss}, \citenamefont {Neudecker}, \citenamefont {Weiss},
  \citenamefont {Back},\ and\ \citenamefont {Schutz}}]{Van-Waeyenberge-2006fk}%
  \BibitemOpen
  \bibfield  {author} {\bibinfo {author} {\bibfnamefont {B.}~\bibnamefont
  {Van~Waeyenberge}}, \bibinfo {author} {\bibfnamefont {A.}~\bibnamefont
  {Puzic}}, \bibinfo {author} {\bibfnamefont {H.}~\bibnamefont {Stoll}},
  \bibinfo {author} {\bibfnamefont {K.~W.}\ \bibnamefont {Chou}}, \bibinfo
  {author} {\bibfnamefont {T.}~\bibnamefont {Tyliszczak}}, \bibinfo {author}
  {\bibfnamefont {R.}~\bibnamefont {Hertel}}, \bibinfo {author} {\bibfnamefont
  {M.}~\bibnamefont {Fahnle}}, \bibinfo {author} {\bibfnamefont
  {H.}~\bibnamefont {Bruckl}}, \bibinfo {author} {\bibfnamefont
  {K.}~\bibnamefont {Rott}}, \bibinfo {author} {\bibfnamefont {G.}~\bibnamefont
  {Reiss}}, \bibinfo {author} {\bibfnamefont {I.}~\bibnamefont {Neudecker}},
  \bibinfo {author} {\bibfnamefont {D.}~\bibnamefont {Weiss}}, \bibinfo
  {author} {\bibfnamefont {C.~H.}\ \bibnamefont {Back}}, \ and\ \bibinfo
  {author} {\bibfnamefont {G.}~\bibnamefont {Schutz}},\ }\href
  {http://dx.doi.org/10.1038/nature05240} {\bibfield  {journal} {\bibinfo
  {journal} {Nature}\ }\textbf {\bibinfo {volume} {444}},\ \bibinfo {pages}
  {461} (\bibinfo {year} {2006})}\BibitemShut {NoStop}%
\bibitem [{\citenamefont {Yamada}\ \emph {et~al.}(2008)\citenamefont {Yamada},
  \citenamefont {Kasai}, \citenamefont {Nakatani}, \citenamefont {Kobayashi},\
  and\ \citenamefont {Ono}}]{Yamada-2008eu}%
  \BibitemOpen
  \bibfield  {author} {\bibinfo {author} {\bibfnamefont {K.}~\bibnamefont
  {Yamada}}, \bibinfo {author} {\bibfnamefont {S.}~\bibnamefont {Kasai}},
  \bibinfo {author} {\bibfnamefont {Y.}~\bibnamefont {Nakatani}}, \bibinfo
  {author} {\bibfnamefont {K.}~\bibnamefont {Kobayashi}}, \ and\ \bibinfo
  {author} {\bibfnamefont {T.}~\bibnamefont {Ono}},\ }\href
  {http://link.aip.org/link/?APL/93/152502/1} {\bibfield  {journal} {\bibinfo
  {journal} {Appl. Phys. Lett.}\ }\textbf {\bibinfo {volume} {93}},\ \bibinfo
  {pages} {152502} (\bibinfo {year} {2008})}\BibitemShut {NoStop}%
\bibitem [{\citenamefont {Vahaplar}\ \emph {et~al.}(2009)\citenamefont
  {Vahaplar}, \citenamefont {Kalashnikova}, \citenamefont {Kimel},
  \citenamefont {Hinzke}, \citenamefont {Nowak}, \citenamefont {Chantrell},
  \citenamefont {Tsukamoto}, \citenamefont {Itoh}, \citenamefont {Kirilyuk},\
  and\ \citenamefont {Rasing}}]{Vahaplar-2009ys}%
  \BibitemOpen
  \bibfield  {author} {\bibinfo {author} {\bibfnamefont {K.}~\bibnamefont
  {Vahaplar}}, \bibinfo {author} {\bibfnamefont {A.~M.}\ \bibnamefont
  {Kalashnikova}}, \bibinfo {author} {\bibfnamefont {A.~V.}\ \bibnamefont
  {Kimel}}, \bibinfo {author} {\bibfnamefont {D.}~\bibnamefont {Hinzke}},
  \bibinfo {author} {\bibfnamefont {U.}~\bibnamefont {Nowak}}, \bibinfo
  {author} {\bibfnamefont {R.}~\bibnamefont {Chantrell}}, \bibinfo {author}
  {\bibfnamefont {A.}~\bibnamefont {Tsukamoto}}, \bibinfo {author}
  {\bibfnamefont {A.}~\bibnamefont {Itoh}}, \bibinfo {author} {\bibfnamefont
  {A.}~\bibnamefont {Kirilyuk}}, \ and\ \bibinfo {author} {\bibfnamefont
  {T.}~\bibnamefont {Rasing}},\ }\href {\doibase
  10.1103/PhysRevLett.103.117201} {\bibfield  {journal} {\bibinfo  {journal}
  {Phys. Rev. Lett.}\ }\textbf {\bibinfo {volume} {103}},\ \bibinfo {eid}
  {117201} (\bibinfo {year} {2009})}\BibitemShut {NoStop}%
\bibitem [{\citenamefont {Hertel}(2009)}]{Hertel-2009vn}%
  \BibitemOpen
  \bibfield  {author} {\bibinfo {author} {\bibfnamefont {R.}~\bibnamefont
  {Hertel}},\ }\href {\doibase 10.1103/Physics.2.73} {\bibfield  {journal}
  {\bibinfo  {journal} {Physics}\ }\textbf {\bibinfo {volume} {2}},\ \bibinfo
  {eid} {73} (\bibinfo {year} {2009})}\BibitemShut {NoStop}%
\bibitem [{\citenamefont {Taguchi}\ \emph {et~al.}(2012)\citenamefont
  {Taguchi}, \citenamefont {Ohe},\ and\ \citenamefont
  {Tatara}}]{Taguchi-2012uq}%
  \BibitemOpen
  \bibfield  {author} {\bibinfo {author} {\bibfnamefont {K.}~\bibnamefont
  {Taguchi}}, \bibinfo {author} {\bibfnamefont {J.-i.}\ \bibnamefont {Ohe}}, \
  and\ \bibinfo {author} {\bibfnamefont {G.}~\bibnamefont {Tatara}},\ }\href
  {\doibase 10.1103/PhysRevLett.109.127204} {\bibfield  {journal} {\bibinfo
  {journal} {Phys. Rev. Lett.}\ }\textbf {\bibinfo {volume} {109}},\ \bibinfo
  {pages} {127204} (\bibinfo {year} {2012})}\BibitemShut {NoStop}%
\bibitem [{\citenamefont {Landau}\ and\ \citenamefont
  {Lifshitz}(1935)}]{Landau-1935fk}%
  \BibitemOpen
  \bibfield  {author} {\bibinfo {author} {\bibfnamefont {L.}~\bibnamefont
  {Landau}}\ and\ \bibinfo {author} {\bibfnamefont {E.}~\bibnamefont
  {Lifshitz}},\ }\href@noop {} {\bibfield  {journal} {\bibinfo  {journal}
  {Phys. Zeitsch. der Sow.}\ }\textbf {\bibinfo {volume} {8}},\ \bibinfo
  {pages} {153} (\bibinfo {year} {1935})},\ \bibinfo {note} {reprint Ukr. J.
  Phys. Vol. 53, p. 14, 2008}\BibitemShut {NoStop}%
\bibitem [{\citenamefont {Papanicolaou}\ and\ \citenamefont
  {Tomaras}(1991)}]{Papanicolaou-1991it}%
  \BibitemOpen
  \bibfield  {author} {\bibinfo {author} {\bibfnamefont {N.}~\bibnamefont
  {Papanicolaou}}\ and\ \bibinfo {author} {\bibfnamefont {T.~N.}\ \bibnamefont
  {Tomaras}},\ }\href@noop {} {\bibfield  {journal} {\bibinfo  {journal}
  {Nuclear Physics B}\ }\textbf {\bibinfo {volume} {360}},\ \bibinfo {pages}
  {425} (\bibinfo {year} {1991})}\BibitemShut {NoStop}%
\bibitem [{\citenamefont {Thiaville}\ \emph {et~al.}(2003)\citenamefont
  {Thiaville}, \citenamefont {Garc\'{\i}a}, \citenamefont {Dittrich},
  \citenamefont {Miltat},\ and\ \citenamefont {Schrefl}}]{Thiaville-2003rq}%
  \BibitemOpen
  \bibfield  {author} {\bibinfo {author} {\bibfnamefont {A.}~\bibnamefont
  {Thiaville}}, \bibinfo {author} {\bibfnamefont {J.~M.}\ \bibnamefont
  {Garc\'{\i}a}}, \bibinfo {author} {\bibfnamefont {R.}~\bibnamefont
  {Dittrich}}, \bibinfo {author} {\bibfnamefont {J.}~\bibnamefont {Miltat}}, \
  and\ \bibinfo {author} {\bibfnamefont {T.}~\bibnamefont {Schrefl}},\ }\href
  {http://link.aps.org/abstract/PRB/v67/e094410} {\bibfield  {journal}
  {\bibinfo  {journal} {Phys. Rev. B}\ }\textbf {\bibinfo {volume} {67}},\
  \bibinfo {pages} {094410} (\bibinfo {year} {2003})}\BibitemShut {NoStop}%
\bibitem [{\citenamefont {Hertel}\ and\ \citenamefont
  {Schneider}(2006)}]{Hertel-2006ib}%
  \BibitemOpen
  \bibfield  {author} {\bibinfo {author} {\bibfnamefont {R.}~\bibnamefont
  {Hertel}}\ and\ \bibinfo {author} {\bibfnamefont {C.~M.}\ \bibnamefont
  {Schneider}},\ }\href@noop {} {\bibfield  {journal} {\bibinfo  {journal}
  {Phys. Rev. Lett.}\ }\textbf {\bibinfo {volume} {97}},\ \bibinfo {pages}
  {177202} (\bibinfo {year} {2006})}\BibitemShut {NoStop}%
\bibitem [{\citenamefont {Yamada}\ \emph {et~al.}(2007)\citenamefont {Yamada},
  \citenamefont {Kasai}, \citenamefont {Nakatani}, \citenamefont {Kobayashi},
  \citenamefont {Kohno}, \citenamefont {Thiaville},\ and\ \citenamefont
  {Ono}}]{Yamada-2007qt}%
  \BibitemOpen
  \bibfield  {author} {\bibinfo {author} {\bibfnamefont {K.}~\bibnamefont
  {Yamada}}, \bibinfo {author} {\bibfnamefont {S.}~\bibnamefont {Kasai}},
  \bibinfo {author} {\bibfnamefont {Y.}~\bibnamefont {Nakatani}}, \bibinfo
  {author} {\bibfnamefont {K.}~\bibnamefont {Kobayashi}}, \bibinfo {author}
  {\bibfnamefont {H.}~\bibnamefont {Kohno}}, \bibinfo {author} {\bibfnamefont
  {A.}~\bibnamefont {Thiaville}}, \ and\ \bibinfo {author} {\bibfnamefont
  {T.}~\bibnamefont {Ono}},\ }\href@noop {} {\bibfield  {journal} {\bibinfo
  {journal} {Nature Mater.}\ }\textbf {\bibinfo {volume} {6}},\ \bibinfo
  {pages} {269} (\bibinfo {year} {2007})}\BibitemShut {NoStop}%
\bibitem [{\citenamefont {Guslienko}\ \emph {et~al.}(2008)\citenamefont
  {Guslienko}, \citenamefont {Lee},\ and\ \citenamefont
  {Kim}}]{Guslienko-2008so}%
  \BibitemOpen
  \bibfield  {author} {\bibinfo {author} {\bibfnamefont {K.~Y.}\ \bibnamefont
  {Guslienko}}, \bibinfo {author} {\bibfnamefont {K.~S.}\ \bibnamefont {Lee}},
  \ and\ \bibinfo {author} {\bibfnamefont {S.~K.}\ \bibnamefont {Kim}},\
  }\href@noop {} {\bibfield  {journal} {\bibinfo  {journal} {Phys. Rev. Lett.}\
  }\textbf {\bibinfo {volume} {100}},\ \bibinfo {pages} {027203} (\bibinfo
  {year} {2008})}\BibitemShut {NoStop}%
\bibitem [{\citenamefont {Gaididei}\ \emph {et~al.}(2008)\citenamefont
  {Gaididei}, \citenamefont {Kravchuk}, \citenamefont {Sheka},\ and\
  \citenamefont {Mertens}}]{Gaididei-2008vn}%
  \BibitemOpen
  \bibfield  {author} {\bibinfo {author} {\bibfnamefont {Y.~B.}\ \bibnamefont
  {Gaididei}}, \bibinfo {author} {\bibfnamefont {V.~P.}\ \bibnamefont
  {Kravchuk}}, \bibinfo {author} {\bibfnamefont {D.~D.}\ \bibnamefont {Sheka}},
  \ and\ \bibinfo {author} {\bibfnamefont {F.~G.}\ \bibnamefont {Mertens}},\
  }\href {http://link.aip.org/link/?LTP/34/528/1} {\bibfield  {journal}
  {\bibinfo  {journal} {Low Temp. Phys.}\ }\textbf {\bibinfo {volume} {34}},\
  \bibinfo {pages} {528} (\bibinfo {year} {2008})}\BibitemShut {NoStop}%
\bibitem [{\citenamefont {Weigand}\ \emph {et~al.}(2009)\citenamefont
  {Weigand}, \citenamefont {Van~Waeyenberge}, \citenamefont {Vansteenkiste},
  \citenamefont {Curcic}, \citenamefont {Sackmann}, \citenamefont {Stoll},
  \citenamefont {Tyliszczak}, \citenamefont {Kaznatcheev}, \citenamefont
  {Bertwistle}, \citenamefont {Woltersdorf}, \citenamefont {Back},\ and\
  \citenamefont {Schutz}}]{Weigand-2009la}%
  \BibitemOpen
  \bibfield  {author} {\bibinfo {author} {\bibfnamefont {M.}~\bibnamefont
  {Weigand}}, \bibinfo {author} {\bibfnamefont {B.}~\bibnamefont
  {Van~Waeyenberge}}, \bibinfo {author} {\bibfnamefont {A.}~\bibnamefont
  {Vansteenkiste}}, \bibinfo {author} {\bibfnamefont {M.}~\bibnamefont
  {Curcic}}, \bibinfo {author} {\bibfnamefont {V.}~\bibnamefont {Sackmann}},
  \bibinfo {author} {\bibfnamefont {H.}~\bibnamefont {Stoll}}, \bibinfo
  {author} {\bibfnamefont {T.}~\bibnamefont {Tyliszczak}}, \bibinfo {author}
  {\bibfnamefont {K.}~\bibnamefont {Kaznatcheev}}, \bibinfo {author}
  {\bibfnamefont {D.}~\bibnamefont {Bertwistle}}, \bibinfo {author}
  {\bibfnamefont {G.}~\bibnamefont {Woltersdorf}}, \bibinfo {author}
  {\bibfnamefont {C.~H.}\ \bibnamefont {Back}}, \ and\ \bibinfo {author}
  {\bibfnamefont {G.}~\bibnamefont {Schutz}},\ }\href
  {http://link.aps.org/abstract/PRL/v102/e077201} {\bibfield  {journal}
  {\bibinfo  {journal} {Phys. Rev. Lett.}\ }\textbf {\bibinfo {volume} {102}},\
  \bibinfo {pages} {077201} (\bibinfo {year} {2009})}\BibitemShut {NoStop}%
\bibitem [{\citenamefont {Bazaliy}\ \emph {et~al.}(1998)\citenamefont
  {Bazaliy}, \citenamefont {Jones},\ and\ \citenamefont
  {Zhang}}]{Bazaliy-1998qf}%
  \BibitemOpen
  \bibfield  {author} {\bibinfo {author} {\bibfnamefont {Y.~B.}\ \bibnamefont
  {Bazaliy}}, \bibinfo {author} {\bibfnamefont {B.~A.}\ \bibnamefont {Jones}},
  \ and\ \bibinfo {author} {\bibfnamefont {S.-C.}\ \bibnamefont {Zhang}},\
  }\href {http://link.aps.org/abstract/PRB/v57/pR3213} {\bibfield  {journal}
  {\bibinfo  {journal} {Physical Review B}\ }\textbf {\bibinfo {volume} {57}}
  (\bibinfo {year} {1998})}\BibitemShut {NoStop}%
\bibitem [{\citenamefont {Zhang}\ and\ \citenamefont
  {Li}(2004)}]{Zhang-2004ve}%
  \BibitemOpen
  \bibfield  {author} {\bibinfo {author} {\bibfnamefont {S.}~\bibnamefont
  {Zhang}}\ and\ \bibinfo {author} {\bibfnamefont {Z.}~\bibnamefont {Li}},\
  }\href {http://link.aps.org/abstract/PRL/v93/e127204} {\bibfield  {journal}
  {\bibinfo  {journal} {Phys. Rev. Lett.}\ }\textbf {\bibinfo {volume} {93}},\
  \bibinfo {pages} {127204} (\bibinfo {year} {2004})}\BibitemShut {NoStop}%
\bibitem [{\citenamefont {Miltat}\ and\ \citenamefont
  {Thiaville}(2002)}]{Miltat-2002kx}%
  \BibitemOpen
  \bibfield  {author} {\bibinfo {author} {\bibfnamefont {J.}~\bibnamefont
  {Miltat}}\ and\ \bibinfo {author} {\bibfnamefont {A.}~\bibnamefont
  {Thiaville}},\ }\href {http://www.sciencemag.org} {\bibfield  {journal}
  {\bibinfo  {journal} {Science}\ }\textbf {\bibinfo {volume} {298}},\ \bibinfo
  {pages} {555} (\bibinfo {year} {2002})}\BibitemShut {NoStop}%
\bibitem [{\citenamefont {Tatara}\ \emph {et~al.}(2008)\citenamefont {Tatara},
  \citenamefont {Kohno},\ and\ \citenamefont {Shibata}}]{Tatara-2008kx}%
  \BibitemOpen
  \bibfield  {author} {\bibinfo {author} {\bibfnamefont {G.}~\bibnamefont
  {Tatara}}, \bibinfo {author} {\bibfnamefont {H.}~\bibnamefont {Kohno}}, \
  and\ \bibinfo {author} {\bibfnamefont {J.}~\bibnamefont {Shibata}},\ }\href
  {http://www.sciencedirect.com/science/article/B6TVP-4T9VP4P-1/2/1d39252a95415cb4df04a52a7c6a755f}
  {\bibfield  {journal} {\bibinfo  {journal} {Phys. Rep.}\ }\textbf {\bibinfo
  {volume} {468}},\ \bibinfo {pages} {213} (\bibinfo {year}
  {2008})}\BibitemShut {NoStop}%
\bibitem [{\citenamefont {Thiaville}\ \emph {et~al.}(2005)\citenamefont
  {Thiaville}, \citenamefont {Nakatani}, \citenamefont {Miltat},\ and\
  \citenamefont {Suzuki}}]{Thiaville-2005cr}%
  \BibitemOpen
  \bibfield  {author} {\bibinfo {author} {\bibfnamefont {A.}~\bibnamefont
  {Thiaville}}, \bibinfo {author} {\bibfnamefont {Y.}~\bibnamefont {Nakatani}},
  \bibinfo {author} {\bibfnamefont {J.}~\bibnamefont {Miltat}}, \ and\ \bibinfo
  {author} {\bibfnamefont {Y.}~\bibnamefont {Suzuki}},\ }\href@noop {}
  {\bibfield  {journal} {\bibinfo  {journal} {EPL (Europhysics Letters)}\
  }\textbf {\bibinfo {volume} {69}},\ \bibinfo {pages} {990} (\bibinfo {year}
  {2005})}\BibitemShut {NoStop}%
\bibitem [{\citenamefont {Ohe}\ and\ \citenamefont
  {Kramer}(2006)}]{Ohe-2006db}%
  \BibitemOpen
  \bibfield  {author} {\bibinfo {author} {\bibfnamefont {J.-i.}\ \bibnamefont
  {Ohe}}\ and\ \bibinfo {author} {\bibfnamefont {B.}~\bibnamefont {Kramer}},\
  }\href {http://link.aps.org/abstract/PRB/v74/e201305} {\bibfield  {journal}
  {\bibinfo  {journal} {Phys. Rev. B}\ }\textbf {\bibinfo {volume} {74}},\
  \bibinfo {pages} {201305} (\bibinfo {year} {2006})}\BibitemShut {NoStop}%
\bibitem [{\citenamefont {Piechon}\ and\ \citenamefont
  {Thiaville}(2007)}]{Piechon-2007rm}%
  \BibitemOpen
  \bibfield  {author} {\bibinfo {author} {\bibfnamefont {F.}~\bibnamefont
  {Piechon}}\ and\ \bibinfo {author} {\bibfnamefont {A.}~\bibnamefont
  {Thiaville}},\ }\href {http://link.aps.org/abstract/PRB/v75/e174414}
  {\bibfield  {journal} {\bibinfo  {journal} {Physical Review B (Condensed
  Matter and Materials Physics)}\ }\textbf {\bibinfo {volume} {75}},\ \bibinfo
  {pages} {174414} (\bibinfo {year} {2007})}\BibitemShut {NoStop}%
\bibitem [{Note1()}]{Note1}%
  \BibitemOpen
  \bibinfo {note} {See Supplemental Material at [URL] for movies showing the
  time evolution of the skyrmions and vortices driven by the spin-polarized
  current}\BibitemShut {NoStop}%
\bibitem [{\citenamefont {Hertel}\ \emph {et~al.}(2007)\citenamefont {Hertel},
  \citenamefont {Gliga}, \citenamefont {Fahnle},\ and\ \citenamefont
  {Schneider}}]{Hertel-2007jy}%
  \BibitemOpen
  \bibfield  {author} {\bibinfo {author} {\bibfnamefont {R.}~\bibnamefont
  {Hertel}}, \bibinfo {author} {\bibfnamefont {S.}~\bibnamefont {Gliga}},
  \bibinfo {author} {\bibfnamefont {M.}~\bibnamefont {Fahnle}}, \ and\ \bibinfo
  {author} {\bibfnamefont {C.~M.}\ \bibnamefont {Schneider}},\ }\href@noop {}
  {\bibfield  {journal} {\bibinfo  {journal} {Phys. Rev. Lett.}\ }\textbf
  {\bibinfo {volume} {98}},\ \bibinfo {pages} {117201} (\bibinfo {year}
  {2007})}\BibitemShut {NoStop}%
\bibitem [{\citenamefont {Kravchuk}\ \emph {et~al.}(2009)\citenamefont
  {Kravchuk}, \citenamefont {Gaididei},\ and\ \citenamefont
  {Sheka}}]{Kravchuk-2009vn}%
  \BibitemOpen
  \bibfield  {author} {\bibinfo {author} {\bibfnamefont {V.~P.}\ \bibnamefont
  {Kravchuk}}, \bibinfo {author} {\bibfnamefont {Y.}~\bibnamefont {Gaididei}},
  \ and\ \bibinfo {author} {\bibfnamefont {D.~D.}\ \bibnamefont {Sheka}},\
  }\href {http://link.aps.org/abstract/PRB/v80/e100405} {\bibfield  {journal}
  {\bibinfo  {journal} {Phys. Rev. B}\ }\textbf {\bibinfo {volume} {80}},\
  \bibinfo {pages} {100405} (\bibinfo {year} {2009})}\BibitemShut {NoStop}%
\bibitem [{\citenamefont {Kammerer}\ \emph {et~al.}(2011)\citenamefont
  {Kammerer}, \citenamefont {Weigand}, \citenamefont {Curcic}, \citenamefont
  {Noske}, \citenamefont {Sproll}, \citenamefont {Vansteenkiste}, \citenamefont
  {Van~Waeyenberge}, \citenamefont {Stoll}, \citenamefont {Woltersdorf},
  \citenamefont {Back},\ and\ \citenamefont {Schuetz}}]{Kammerer-2011fk}%
  \BibitemOpen
  \bibfield  {author} {\bibinfo {author} {\bibfnamefont {M.}~\bibnamefont
  {Kammerer}}, \bibinfo {author} {\bibfnamefont {M.}~\bibnamefont {Weigand}},
  \bibinfo {author} {\bibfnamefont {M.}~\bibnamefont {Curcic}}, \bibinfo
  {author} {\bibfnamefont {M.}~\bibnamefont {Noske}}, \bibinfo {author}
  {\bibfnamefont {M.}~\bibnamefont {Sproll}}, \bibinfo {author} {\bibfnamefont
  {A.}~\bibnamefont {Vansteenkiste}}, \bibinfo {author} {\bibfnamefont
  {B.}~\bibnamefont {Van~Waeyenberge}}, \bibinfo {author} {\bibfnamefont
  {H.}~\bibnamefont {Stoll}}, \bibinfo {author} {\bibfnamefont
  {G.}~\bibnamefont {Woltersdorf}}, \bibinfo {author} {\bibfnamefont {C.~H.}\
  \bibnamefont {Back}}, \ and\ \bibinfo {author} {\bibfnamefont
  {G.}~\bibnamefont {Schuetz}},\ }\href {http://dx.doi.org/10.1038/ncomms1277}
  {\bibfield  {journal} {\bibinfo  {journal} {Nat Commun}\ }\textbf {\bibinfo
  {volume} {2}},\ \bibinfo {pages} {279} (\bibinfo {year} {2011})}\BibitemShut
  {NoStop}%
\bibitem [{\citenamefont {Zang}\ \emph {et~al.}(2011)\citenamefont {Zang},
  \citenamefont {Mostovoy}, \citenamefont {Han},\ and\ \citenamefont
  {Nagaosa}}]{Zang-2011fk}%
  \BibitemOpen
  \bibfield  {author} {\bibinfo {author} {\bibfnamefont {J.}~\bibnamefont
  {Zang}}, \bibinfo {author} {\bibfnamefont {M.}~\bibnamefont {Mostovoy}},
  \bibinfo {author} {\bibfnamefont {J.~H.}\ \bibnamefont {Han}}, \ and\
  \bibinfo {author} {\bibfnamefont {N.}~\bibnamefont {Nagaosa}},\ }\href
  {\doibase 10.1103/PhysRevLett.107.136804} {\bibfield  {journal} {\bibinfo
  {journal} {Phys. Rev. Lett.}\ }\textbf {\bibinfo {volume} {107}},\ \bibinfo
  {pages} {136804} (\bibinfo {year} {2011})}\BibitemShut {NoStop}%
\bibitem [{\citenamefont {Berger}(1984)}]{Berger-1984uq}%
  \BibitemOpen
  \bibfield  {author} {\bibinfo {author} {\bibfnamefont {L.}~\bibnamefont
  {Berger}},\ }\href
  {http://scitation.aip.org/content/aip/journal/jap/55/6/10.1063/1.333530}
  {\bibfield  {journal} {\bibinfo  {journal} {J. Appl. Phys.}\ }\textbf
  {\bibinfo {volume} {55}},\ \bibinfo {pages} {1954} (\bibinfo {year}
  {1984})}\BibitemShut {NoStop}%
\bibitem [{\citenamefont {Grigoriev}\ \emph {et~al.}(2006)\citenamefont
  {Grigoriev}, \citenamefont {Maleyev}, \citenamefont {Okorokov}, \citenamefont
  {Chetverikov}, \citenamefont {B\"oni}, \citenamefont {Georgii}, \citenamefont
  {Lamago}, \citenamefont {Eckerlebe},\ and\ \citenamefont
  {Pranzas}}]{Grigoriev-2006ys}%
  \BibitemOpen
  \bibfield  {author} {\bibinfo {author} {\bibfnamefont {S.~V.}\ \bibnamefont
  {Grigoriev}}, \bibinfo {author} {\bibfnamefont {S.~V.}\ \bibnamefont
  {Maleyev}}, \bibinfo {author} {\bibfnamefont {A.~I.}\ \bibnamefont
  {Okorokov}}, \bibinfo {author} {\bibfnamefont {Y.~O.}\ \bibnamefont
  {Chetverikov}}, \bibinfo {author} {\bibfnamefont {P.}~\bibnamefont {B\"oni}},
  \bibinfo {author} {\bibfnamefont {R.}~\bibnamefont {Georgii}}, \bibinfo
  {author} {\bibfnamefont {D.}~\bibnamefont {Lamago}}, \bibinfo {author}
  {\bibfnamefont {H.}~\bibnamefont {Eckerlebe}}, \ and\ \bibinfo {author}
  {\bibfnamefont {K.}~\bibnamefont {Pranzas}},\ }\href {\doibase
  10.1103/PhysRevB.74.214414} {\bibfield  {journal} {\bibinfo  {journal} {Phys.
  Rev. B}\ }\textbf {\bibinfo {volume} {74}},\ \bibinfo {pages} {214414}
  (\bibinfo {year} {2006})}\BibitemShut {NoStop}%
\bibitem [{\citenamefont {St{\"o}hr}\ and\ \citenamefont
  {Siegmann}(2006)}]{Stohr-2006qf}%
  \BibitemOpen
  \bibfield  {author} {\bibinfo {author} {\bibfnamefont {J.}~\bibnamefont
  {St{\"o}hr}}\ and\ \bibinfo {author} {\bibfnamefont {H.}~\bibnamefont
  {Siegmann}},\ }\href {http://books.google.fr/books?id=oENBAQAAIAAJ} {\emph
  {\bibinfo {title} {Magnetism: from Fundamentals to Nanoscale Dynamics}}},\
  Springer series in solid-state sciences\ (\bibinfo  {publisher} {Springer},\
  \bibinfo {year} {2006})\BibitemShut {NoStop}%
\bibitem [{\citenamefont {Liu}\ \emph {et~al.}(2007)\citenamefont {Liu},
  \citenamefont {Gliga}, \citenamefont {Hertel},\ and\ \citenamefont
  {Schneider}}]{Liu-2007hw}%
  \BibitemOpen
  \bibfield  {author} {\bibinfo {author} {\bibfnamefont {Y.}~\bibnamefont
  {Liu}}, \bibinfo {author} {\bibfnamefont {S.}~\bibnamefont {Gliga}}, \bibinfo
  {author} {\bibfnamefont {R.}~\bibnamefont {Hertel}}, \ and\ \bibinfo {author}
  {\bibfnamefont {C.~M.}\ \bibnamefont {Schneider}},\ }\href
  {http://link.aip.org/link/?APL/91/112501/1} {\bibfield  {journal} {\bibinfo
  {journal} {Applied Physics Letters}\ }\textbf {\bibinfo {volume} {91}},\
  \bibinfo {pages} {112501} (\bibinfo {year} {2007})}\BibitemShut {NoStop}%
\bibitem [{\citenamefont {Nakatani}\ \emph {et~al.}(2008)\citenamefont
  {Nakatani}, \citenamefont {Shibata}, \citenamefont {Tatara}, \citenamefont
  {Kohno}, \citenamefont {Thiaville},\ and\ \citenamefont
  {Miltat}}]{Nakatani-2008pz}%
  \BibitemOpen
  \bibfield  {author} {\bibinfo {author} {\bibfnamefont {Y.}~\bibnamefont
  {Nakatani}}, \bibinfo {author} {\bibfnamefont {J.}~\bibnamefont {Shibata}},
  \bibinfo {author} {\bibfnamefont {G.}~\bibnamefont {Tatara}}, \bibinfo
  {author} {\bibfnamefont {H.}~\bibnamefont {Kohno}}, \bibinfo {author}
  {\bibfnamefont {A.}~\bibnamefont {Thiaville}}, \ and\ \bibinfo {author}
  {\bibfnamefont {J.}~\bibnamefont {Miltat}},\ }\href@noop {} {\bibfield
  {journal} {\bibinfo  {journal} {Phys. Rev. B}\ }\textbf {\bibinfo {volume}
  {77}},\ \bibinfo {pages} {014439} (\bibinfo {year} {2008})}\BibitemShut
  {NoStop}%
\bibitem [{\citenamefont {Tchoe}\ and\ \citenamefont
  {Han}(2012)}]{Tchoe-2012uq}%
  \BibitemOpen
  \bibfield  {author} {\bibinfo {author} {\bibfnamefont {Y.}~\bibnamefont
  {Tchoe}}\ and\ \bibinfo {author} {\bibfnamefont {J.~H.}\ \bibnamefont
  {Han}},\ }\href {\doibase 10.1103/PhysRevB.85.174416} {\bibfield  {journal}
  {\bibinfo  {journal} {Phys. Rev. B}\ }\textbf {\bibinfo {volume} {85}},\
  \bibinfo {pages} {174416} (\bibinfo {year} {2012})}\BibitemShut {NoStop}%
\bibitem [{\citenamefont {{Kim}}\ and\ \citenamefont
  {{Onoda}}(2010)}]{Kim-2010uq}%
  \BibitemOpen
  \bibfield  {author} {\bibinfo {author} {\bibfnamefont {K.-S.}\ \bibnamefont
  {{Kim}}}\ and\ \bibinfo {author} {\bibfnamefont {S.}~\bibnamefont
  {{Onoda}}},\ }\href@noop {} {\bibfield  {journal} {\bibinfo  {journal} {ArXiv
  e-prints}\ } (\bibinfo {year} {2010})},\ \Eprint
  {http://arxiv.org/abs/1012.0631} {arXiv:1012.0631 [cond-mat.mes-hall]}
  \BibitemShut {NoStop}%
\bibitem [{\citenamefont {Jonietz}\ \emph {et~al.}(2010)\citenamefont
  {Jonietz}, \citenamefont {M{\"u}hlbauer}, \citenamefont {Pfleiderer},
  \citenamefont {Neubauer}, \citenamefont {M{\"u}nzer}, \citenamefont {Bauer},
  \citenamefont {Adams}, \citenamefont {Georgii}, \citenamefont {B{\"o}ni},
  \citenamefont {Duine}, \citenamefont {Everschor}, \citenamefont {Garst},\
  and\ \citenamefont {Rosch}}]{Jonietz-2010ly}%
  \BibitemOpen
  \bibfield  {author} {\bibinfo {author} {\bibfnamefont {F.}~\bibnamefont
  {Jonietz}}, \bibinfo {author} {\bibfnamefont {S.}~\bibnamefont
  {M{\"u}hlbauer}}, \bibinfo {author} {\bibfnamefont {C.}~\bibnamefont
  {Pfleiderer}}, \bibinfo {author} {\bibfnamefont {A.}~\bibnamefont
  {Neubauer}}, \bibinfo {author} {\bibfnamefont {W.}~\bibnamefont
  {M{\"u}nzer}}, \bibinfo {author} {\bibfnamefont {A.}~\bibnamefont {Bauer}},
  \bibinfo {author} {\bibfnamefont {T.}~\bibnamefont {Adams}}, \bibinfo
  {author} {\bibfnamefont {R.}~\bibnamefont {Georgii}}, \bibinfo {author}
  {\bibfnamefont {P.}~\bibnamefont {B{\"o}ni}}, \bibinfo {author}
  {\bibfnamefont {R.~A.}\ \bibnamefont {Duine}}, \bibinfo {author}
  {\bibfnamefont {K.}~\bibnamefont {Everschor}}, \bibinfo {author}
  {\bibfnamefont {M.}~\bibnamefont {Garst}}, \ and\ \bibinfo {author}
  {\bibfnamefont {A.}~\bibnamefont {Rosch}},\ }\href@noop {} {\bibfield
  {journal} {\bibinfo  {journal} {Science}\ }\textbf {\bibinfo {volume}
  {330}},\ \bibinfo {pages} {1648} (\bibinfo {year} {2010})}\BibitemShut
  {NoStop}%
\bibitem [{\citenamefont {Romming}\ \emph {et~al.}(2013)\citenamefont
  {Romming}, \citenamefont {Hanneken}, \citenamefont {Menzel}, \citenamefont
  {Bickel}, \citenamefont {Wolter}, \citenamefont {von Bergmann}, \citenamefont
  {Kubetzka},\ and\ \citenamefont {Wiesendanger}}]{Romming-2013lq}%
  \BibitemOpen
  \bibfield  {author} {\bibinfo {author} {\bibfnamefont {N.}~\bibnamefont
  {Romming}}, \bibinfo {author} {\bibfnamefont {C.}~\bibnamefont {Hanneken}},
  \bibinfo {author} {\bibfnamefont {M.}~\bibnamefont {Menzel}}, \bibinfo
  {author} {\bibfnamefont {J.~E.}\ \bibnamefont {Bickel}}, \bibinfo {author}
  {\bibfnamefont {B.}~\bibnamefont {Wolter}}, \bibinfo {author} {\bibfnamefont
  {K.}~\bibnamefont {von Bergmann}}, \bibinfo {author} {\bibfnamefont
  {A.}~\bibnamefont {Kubetzka}}, \ and\ \bibinfo {author} {\bibfnamefont
  {R.}~\bibnamefont {Wiesendanger}},\ }\href
  {http://www.sciencemag.org/content/341/6146/636.abstract N2 - Topologically
  nontrivial spin textures have recently been investigated for spintronic
  applications. Here, we report on an ultrathin magnetic film in which
  individual skyrmions can be written and deleted in a controlled fashion with
  local spin-polarized currents from a scanning tunneling microscope. An
  external magnetic field is used to tune the energy landscape, and the
  temperature is adjusted to prevent thermally activated switching between
  topologically distinct states. Switching rate and direction can then be
  controlled by the parameters used for current injection. The creation and
  annihilation of individual magnetic skyrmions demonstrates the potential for
  topological charge in future information-storage concepts.} {\bibfield
  {journal} {\bibinfo  {journal} {Science}\ }\textbf {\bibinfo {volume}
  {341}},\ \bibinfo {pages} {636} (\bibinfo {year} {2013})}\BibitemShut
  {NoStop}%
\end{thebibliography}
\end{document}